\documentclass[11pt,reqno]{amsart}
\let\oldll\ll
\let\oldgg\gg
\usepackage{amssymb,mathtools,mathabx,amsthm}
\usepackage{eucal}
\renewcommand{\ll}{\oldll}
\renewcommand{\gg}{\oldgg}

\usepackage[margin=1in]{geometry}

\usepackage{enumerate}

\usepackage{mathrsfs,bm}
\usepackage[usenames,dvipsnames,svgnames,table]{xcolor}
\usepackage[pagebackref,letterpaper=true,colorlinks=true,pdfpagemode=none,citecolor=OliveGreen,linkcolor=RoyalPurple,urlcolor=BrickRed,pdfstartview=FitH]{hyperref}
\usepackage[font=small,justification=centering]{caption}
\usepackage[font=footnotesize,justification=centering]{subcaption}
\usepackage{framed}

\newcommand{\f}{\frac}

\renewcommand{\vec}[1]{\underline{\smash{#1}}}

\newcommand{\beq}{\begin{equation}}
\newcommand{\eeq}{\end{equation}}
\newtheorem{thm}{Theorem}
\newtheorem{lem}{Lemma}
\newtheorem{cor}[lem]{Corollary}
\newtheorem{ppn}[lem]{Proposition}
\theoremstyle{definition}
\newtheorem{ass}{Assumption}
\newtheorem{alg}{Algorithm}
\newtheorem{dfn}[lem]{Definition}
\newtheorem{rmk}[lem]{Remark}
\newcommand{\set}[1]{\{#1\}}

\newcommand{\E}{\mathbb{E}}
\renewcommand{\P}{\mathbb{P}}
\newcommand{\R}{\mathbb{R}}

\newcommand{\Ind}[1]{\mathbf{1}\{#1\}}
\def\<{\langle}
\def\>{\rangle}

\DeclareMathOperator{\diag}{\textup{\textsf{diag}}}
\DeclareMathOperator{\rank}{\textup{\textsf{rank}}}
\DeclareMathOperator{\trace}{\textup{\textsf{tr}}}
\DeclareMathOperator{\polylog}{polylog}
\DeclareMathOperator{\spn}{\textsf{span}}
\DeclareMathOperator{\esssup}{ess\,sup}

\newcommand{\Proj}{\Pi}
\newcommand{\LL}{\lambda}
\newcommand{\RR}{\rho}
\newcommand{\MM}{\gamma}
\newcommand{\ALPHA}{\alpha}
\newcommand{\MU}{\mu}
\newcommand{\THETA}{\vartheta}
\newcommand{\ETA}{\eta}

\newcommand{\uu}{{\vec{u}}}
\newcommand{\ui}{\vec{i}}
\newcommand{\uj}{\vec{j}}

\newcommand{\ul}{\vec{\ell}}
\newcommand{\coher}{\textup{\textsf{coher}}}
\newcommand{\unfold}{\textup{\textsf{unfold}}}

\def\de{{\rm d}}
\newcommand{\bF}{\bm{F}}

\newcommand{\bT}{\bm{T}}
\newcommand{\bV}{\bm{V}}
\newcommand{\bY}{\bm{Y}}
\newcommand{\bTstar}{\bT^\star}
\newcommand{\hbT}{{\widehat{\bm{T}}}}
\newcommand{\hT}{\widehat{T}}
\newcommand{\hbTstar}{\widehat{\bm{T}}^\star}

\def\Bern{{\rm Ber}}

\def\hB{\widehat{B}}

\def\sT{{\sf t}}
\def\id{I}

\def\RU{\textup{\textsc{r}}}
\newcommand\rmu[1]{r_{\textup{\tiny$\boxplus$},#1}}
\def\rmum{r_{\textup{\tiny$\boxplus$},\mbox{\tiny\rm max}}}
\def\op{\textup{op}}
\def\F{\textup{F}}
\def\cQ{{\mathcal{Q}}}

\newcommand{\lmstar}{\lambda_\star}

\newcommand{\sgns}{\mathfrak{s}}
\newcommand{\sgnt}{\mathfrak{t}}

\newcommand{\cross}{\textup{cross}}

\newcommand{\dg}{\textup{diag}}
\newcommand{\sign}{\textup{sign}}

\newcommand{\I}{\mathbf{1}}

\newcommand{\nn}{\textsf{\itshape n}}
\newcommand{\cc}{\textsf{\itshape C}}
\newcommand{\ee}{\textup{\textsf{E}}}
\newcommand{\dd}{\textup{\textsf{D}}}
\newcommand{\II}{\textup{\textsf{I}}}
\newcommand{\JJ}{\textup{\textsf{J}}}
\newcommand{\KK}{\textup{\textsf{K}}}

\title[Spectral algorithms for tensor completion]{Spectral algorithms for tensor completion}
\author[A.~Montanari]{Andrea~Montanari}
\author[N.~Sun]{Nike~Sun}

\begin{document}

\maketitle

\begin{abstract} In the tensor completion problem, one seeks to estimate a low-rank tensor 
based on a random sample of revealed entries. In terms of the required sample size, earlier work revealed a large gap between estimation with unbounded computational resources (using, for instance, tensor nuclear norm minimization) and polynomial-time algorithms. Among the latter, the best statistical guarantees have been proved, for third-order tensors, using the sixth level of the sum-of-squares (\textsc{sos}) semidefinite programming hierarchy (Barak and Moitra, 2014). However, the \textsc{sos} approach does not scale well to large problem instances. By contrast, spectral methods --- based on unfolding or matricizing the tensor --- are attractive for their low complexity, but have been believed to require a much larger sample size.

This paper presents two main contributions. First, we propose a new unfolding-based method, which outperforms naive ones for symmetric $k$-th order tensors of rank $r$. For this result we make a study of singular space estimation for partially revealed matrices of large aspect ratio, which may be of independent interest. For third-order tensors, our algorithm matches the \textsc{sos} method in terms of sample size (requiring about $rd^{3/2}$ revealed entries), subject to a worse rank condition ($r\ll d^{3/4}$ rather than $r\ll d^{3/2}$). We complement this result with a different spectral algorithm for third-order tensors in the overcomplete ($r\ge d$) regime. Under a random model, this second approach succeeds in estimating tensors of rank $d\le r \ll d^{3/2}$ from about $rd^{3/2}$ revealed entries.
\end{abstract}

\section{Introduction} Tensors are increasingly ubiquitous in a variety of statistics and machine learning contexts. Many datasets are arranged according to the values of three or more attributes, giving rise to multi-way tables which can be interpreted as tensors \cite{morup2011applications}. For instance, consider the collaborative filtering problem in which a group of users provide feedback on the episodes of a certain number of television shows, over an extended time interval. The data is indexed by three attributes --- user \textsc{id}, show \textsc{id}, and episode broadcast time --- so it is presented as a three-way table (which is a tensor). A second example comes from high-dimensional applications of the moment method \cite{hsu2012spectral}: the $k$-th moments of a multivariate distribution are naturally encoded by a $k$-fold tensor. Some other applications include image inpainting \cite{liu2013tensor}, hyperspectral imaging \cite{li2010tensor,signoretto2011tensor}, and geophysical imaging \cite{kreimer2013tensor}.

In many applications, the underlying tensor $\bT$ is only partially observed, and it is of interest to use the observed entries to impute the missing ones. This is the \emph{tensor completion problem}. Clearly, completion is plausible only if the underlying tensor $\bT$ is sufficiently structured: it is standard to posit that it has low rank, and is incoherent with respect to standard basis vectors. These assumptions are formalized in a few non-equivalent ways in the existing literature; we review some of these below. We assume
an underlying order-$k$ tensor,
$\bT\in(\R^d)^{\otimes k}$
--- it is a $k$-way array, with entries $\bT_\uu$ indexed by $\uu\in[d]^k$ where $[d]\equiv\set{1,\ldots,d}$. 
Our basic structural assumption is that $\bT$ has low rank in the sense that it 
is expressible as a sum of $r$ pure tensors:
	\beq\label{e:sum.r.pure.tensors}
	\bT = \sum_{s\le r}
		a^{(1)}_s
		\otimes \cdots\otimes
		a^{(k)}_s\,.
	\eeq
This paper proposes methods for completing $\bT$ from $\nn$ observed entries, and investigates
the minimum number $\nn$ (as a function of $k,d,r$) required for a non-trivial estimator.

\subsection{Related work}

There is already a substantial literature on tensor completion, and we survey here some of the main ideas that have emerged.

\subsubsection*{Non-polynomial estimators} Motivated by the success of 
methods for matrix completion
 based on
nuclear norm relaxations  \cite{MR2565240,gross2011recovering}, several papers have studied estimators based on a suitable definition of tensor nuclear norm \cite{yuan2015tensor,yuan2016incoherent}. This tensor norm is  \textsc{np}-hard to evaluate \cite{friedland2016nuclear} 
and therefore this approach does not lead to practical algorithms. Nevertheless these studies provide useful information on the minimum number $\nn$ of entries 
required to reconstruct $\bT$ 
\emph{with unbounded computational resources}. In particular, it was proved
\cite{yuan2016incoherent} that it suffices to have
	\[
	\nn\ge\cc
	d(\log d)^2
	\max\bigg\{
	(\rmum)^{k-1},
	(\rmum)^{(k-1)/2}d^{1/2}
	\bigg\}\,,
	\]
with $\rmum$ the multilinear (or Tucker) rank of $\bT$. Here we use $\cc$ to denote a constant that can depend on various incoherence parameters; in later sections we will make such factors explicit. The definition of $\rmum$ is reviewed below; we comment also that
$r^{1/(k-1)}\le\rmum\le
\min\set{d,r}$ (see \eqref{e:r.ubd.tucker.rank}). 
Information-theoretic considerations also indicate that
	\beq\label{e:information.lbd}
	\nn\ge\cc r d\eeq
entries are necessary --- indeed, the number of parameters required to specify a tensor $\bT\in (\R^d)^{\otimes k}$ of rank $r$ is of order $rd$ (we treat $k$ as a constant throughout).

\subsubsection*{Tensor unfolding}
At the opposite extreme, tensor unfolding gives access to very efficient matrix completion algorithms. For integers $a,b\ge1$ with $a+b=k$, a tensor $\bT\in(\R^d)^{\otimes k}$ can be unfolded into a $d^a\times d^b$ matrix. Formally, the unfolding operation is a linear map
	\[\unfold^{a\times b}
	: (\R^d)^{\otimes k}
	\to \R^{d^a\times d^b},\quad
	\bT\mapsto X\]
where $X_{\ui,\ul}=\bT_\uu$
for $\ui\in[d]^a$, $\ul\in[d]^b$, and 
$\uu=(\ui,\ul)\in[d]^k$. One can then apply matrix completion algorithms --- e.g.\ spectral methods, or convex relaxations --- to $X$, which is sometimes called the $d^a\times d^b$ \emph{matricization} of $\bT$.  Supposing without loss that $a\le b$, results in the matrix completion literature \cite{gross2011recovering,MR2877360} imply exact reconstruction with
	\beq\label{e:spectral.lbd}
	\nn\ge \cc r d^b (\log d)^2\,.
	\eeq
This remark has been applied several times (e.g.\ \cite{tomioka2010estimation, tomioka2011statistical, liu2013tensor, gandy2011tensor}). It seems to suggest two practically important consequences: (i) the unfolding should be  made ``as square as possible'' by taking $a=\lfloor k/2\rfloor$ and $b=\lceil k/2\rceil$ \cite{mu2014square}; and (ii) unfolding-based algorithms are fundamentally limited to a sample size $\nn \ge\cc r d^{\lceil k/2\rceil}$, due to the limitations of matrix completion --- this has been suggested by several authors \cite{yuan2015tensor,yuan2016incoherent,BarakMoitra}, and is further discussed below. One of the main purposes of this paper is to revisit this last insight.

\subsubsection*{Semidefinite programming hierarchies.} 
In terms of the number $\nn$ of observed entries required, the above results indicate a large gap
between  information-theoretic limits
\eqref{e:information.lbd} on the one hand,
and the requirements of spectral algorithms
\eqref{e:spectral.lbd} on the other.
Motivated by this gap,
Barak and Moitra \cite{BarakMoitra}
considered the \emph{sum-of-squares} (\textsc{sos})  hierarchy to design a more powerful polynomial-time algorithm for this problem.

Without entering into the details, the tensor completion problem can be naturally phrased as a polynomial optimization problem, to which the \textsc{sos} framework is particularly suited. It defines a hierarchy of semidefinite programming (\textsc{sdp}) relaxations, indexed by a degree $\ell$ which is a positive even integer. The \mbox{degree-$\ell$} relaxation requires solving a \textsc{sdp} where the decision variable is a $d^{\ell/2}\times d^{\ell/2}$ matrix; this can be done in time $O(d^{5\ell/2})$ by interior-point methods \cite{alizadeh1995interior}. The \textsc{sos} hierarchy is the most powerful \textsc{sdp} hierarchy. It has attracted considerable interest because it matches complexity-theoretic lower bounds in many problems \cite{barak2014sum}.

Barak~and~Moitra consider the completion problem for a tensor $\bT$ of order $k=3$, along with a slightly different notion of rank $r_\star(\bT)$. (It is a relaxation of the tensor nuclear norm of $\bT$, which in turn can be viewed as a relaxation of the rank $r(\bT)$ \cite{friedland2016nuclear}.) The main result of~\cite{BarakMoitra} is that the degree-$6$ level of the \textsc{sos} hierarchy succeeds in completing a tensor of order $k=3$ from
	\beq\label{e:bm.main.lbd}
	\nn\ge\cc 
	(r_\star)^2 d^{3/2}(\log d)^4
	\eeq
entries. Under additional randomness assumptions, it is proved that
	\beq\label{e:bm.random.lbd}
	\nn \ge \cc rd^{3/2} \polylog(d)
	\eeq
entries suffice. Considering the case of bounded rank, the \cite{BarakMoitra} result improves (for $k=3$) over earlier results~\eqref{e:spectral.lbd} obtained by  unfolding, which required $\nn\ge\cc r d^2(\log d)^2$. At the same time it is far from the information-theoretic bound~\eqref{e:information.lbd}, and this remaining gap may be of a fundamental nature: the authors present evidence to suggest that condition~\eqref{e:bm.main.lbd} is nearly-optimal among polynomial-time algorithms.

\subsection{Main contributions}

Let us emphasize that the degree-$6$ \textsc{sos} relaxation requires solving an
\textsc{sdp} for a matrix of dimensions $d^3\times d^3$. This can be done in polynomial time, but practical implementations would hardly scale beyond $d=20$. For this reason we interpret the results of \cite{BarakMoitra} as opening (rather than closing) a search for fast tensor completion algorithms. With this motivation, we present the following results in this paper:

\subsubsection*{Improved unfolding-based estimator} We consider the completion problem for symmetric tensors of general order $k\ge3$, and propose a new estimator which is based on spectral analysis of the unfolded tensor. We show that our estimator
succeeds in completing the tensor given
	\beq\label{e:our.lbd}
	\nn \ge \cc r d^{k/2} \polylog(d)
	\eeq
revealed entries, subject to $r\le r_{\max}(d;k)$ (see~\eqref{e:r.max}). The main input to this result is the following observation. For $d_1\times d_2$ matrices with $d_1\le d_2$, it is well known that completion is impossible, by any means, unless  $\nn\gg rd_2$. (This was noted for example by~\cite{MR2723472} --- consider the $d_1\times d_2$ matrix $X$ whose $i$-th row is given by $v_{\lceil ir/d_1 \rceil }$, for random vectors $v_1,\ldots,v_r\in\R^{d_2}$.) However, we show that the \emph{column} space 
can be estimated with fewer entries, namely
$\smash{\nn\ge r(d_1d_2)^{1/2}
	\polylog(d_2)}$.

Previous unfolding-based methods have essentially performed matrix completion on the unfolded tensor, a $d^a\times d^b$ matrix. As we noted above, if $a\le b$ this necessitates $\nn\gg rd^b$, which is essentially matched by~\eqref{e:spectral.lbd}. By contrast, our algorithm only seeks to estimate the column space of the unfolding, which requires fewer revealed entries, $\nn\gg rd^{(a+b)/2}= rd^{k/2}$. Given our estimate of the singular space, we then take advantage of the original tensor structure to estimate the missing entries.

\subsubsection*{Overcomplete three-tensors} For symmetric tensors of order $k=3$ we can compare our unfolding algorithm with the \textsc{sos} algorithm of \cite{BarakMoitra}. Even with crude methods for matrix operations, the unfolding algorithm takes  at most $O(d^5)$ time, as opposed to $O(d^{15})$ for degree-$6$ \textsc{sos} (using generic \textsc{sdp} solvers). Neglecting logarithmic factors, our result matches theirs in the required sample size (\eqref{e:bm.random.lbd}~versus~\eqref{e:our.lbd}), but with a significantly worse rank condition: we require $r \ll d^{3/4}$ whereas  \textsc{sos}  succeeds up to $r\ll d^{3/2}$. Indeed, for third-order tensors which are overcomplete (rank $r\ge d$), we do not expect that any unfolding-based method can succeed --- the $d\times d^2$ unfolding can have rank at most $d$, and will fail to capture the rank-$r$ tensor structure. Instead, we complement our unfolding algorithm with a more specialized spectral algorithm, which is specifically intended for overcomplete three-tensors; the runtime is $O(d^6)$. In a certain random tensor model, we show that this second method can succesfully estimate three-tensors from $\nn\gg rd^{3/2}$ revealed entries, for $d\le r\ll d^{3/2}$. In the design and analysis of this method we were inspired by some recent work  \cite{hsss} on the tensor decomposition problem.

\subsection{Organization of the paper} In Section~\ref{sec:Notations} we review some definitions and notations. We then state our main results on tensor completion: Section~\ref{sec:Unfold.Tensor} presents the unfolding-based algorithm, and Section~\ref{sec:Overcomplete} presents the more specialized algorithm for overcomplete three-tensors. In Section~\ref{sec:Numerical} we illustrate our results with some numerical simulations. As noted above, for our unfolding algorithm we study the column spaces of partially revealed matrices with large aspect ratio; our results on this are presented in~Section~\ref{sec:Matrix}.

\section{Preliminaries}
\label{sec:Notations}

\subsection{Notation and terminology}
Given two vector spaces $\mathcal{U}$ and $\mathcal{V}$, we let $\mathcal{U}\otimes\mathcal{V}$ denote their tensor product. 
Following standard practice, we frequently identify
$\R^{d_1}\otimes \R^{d_2}$
with $\R^{d_1d_2}$ or with $\R^{d_1\times d_2}$
(the vector space of $d_1\times d_2$ real matrices). We use lower-case letters for scalars
($a,b,c,\ldots$ and Greek letters)
and vectors ($u,v,w,\ldots$).
We use upper-case letters
($A,B,C,\ldots$) for matrices,
and upper-case boldface letters
($\bm{A},\bm{B},\bm{C},\ldots$)
for tensors of order $k\ge3$.
The $d\times d$ identity matrix is denoted by $\id_d$.

Between two tensors (of any order $k\ge1$)
we use $\otimes$ to denote the tensor product.
Between two tensors in the same space we use
$\odot$ to denote the Hadamard (entrywise) product. For instance,
	\[
	(A\odot B)
	\otimes(C\odot D)
	=(A\otimes C)\odot (B\otimes D)\,.
	\]
We use angle brackets $\<\cdot,\cdot\>$ to denote the standard euclidean scalar product
--- regardless of whether the objects involved are
vectors, matrices, or tensors.
For example, if $X,Y$ are two 
$d_1\times d_2$ matrices, then
we use $\<X,Y\>$ to denote the scalar product between $X$ and $Y$  as $(d_1d_2)$-dimensional vectors. The euclidean norm of a vector $v$ will be denoted $\|v\|=\<v,v\>^{1/2}$.
The Frobenius norm of a $d_1\times d_2$ matrix $X$
is $\|X\|_\F=\<X,X\>^{1/2}$; it is the euclidean norm of $X$
regarded as an $(d_1d_2)$-dimensional vector.
Likewise the Frobenius norm of a tensor $\bT$
is $\|\bT\|_\F=\<\bT,\bT\>^{1/2}$.
For a $d_1\times d_2$ matrix $X$
we write $\|X\|_\op$ for its spectral norm (operator norm). Finally, we let $\|X\|_\infty$
denote the maximum entry size of $X$.

For any subset $\ee\subseteq[d_1]\times\cdots\times[d_k]$
we let $\Proj_\ee$ denote the projection on
$\R^{d_1}\otimes \cdots\otimes \R^{d_k}$
which maps $\bT$ to the tensor
$\Proj_\ee\bT$ with entries
	\[(\Proj_\ee\bT)_\uu 
	= \bT_\uu \Ind{\uu\in\ee}\,.\]
In the special case $k=2$ and $d_1=d_2=d$,
we let
	\[\Proj\equiv\Proj_\dd,\quad
	\Proj_\perp \equiv \id_d-\Proj\]
where $\dd
=\set{(i,i) : 1\le i\le d}$ is the set of diagonal entries.

We say that an event $A$ occurs ``with  high probability'' if $\P(A^c)$ tends to zero as the dimension parameter $d=\min\set{d_1,\dots,d_k}$  tends to infinity. We say that $A$ occurs ``with very high probability'' if $\P(A^c)$ tends to zero faster than any polynomial of $d$.  We will frequently take union  bounds over $m$ such events where $m$ is bounded by some polynomial of $d$. For any two functions $f,g$ depending on $(d_1,\ldots,d_k)$, we write $f\lesssim g$ to indicate that $f \le\cc (\log d)^\beta g$ whenever $d=\min\set{d_1,\ldots,d_k}\ge \beta$, where (as before) $\cc$ is a constant which can depend on incoherence parameters, and $\beta$ is a constant which can depend on $k$.

Our main results for tensor completion assume a symmetric underlying tensor $\bT\in(\R^d)^{\otimes k}$. It has entries $\bT_\uu$ indexed by $\uu\in[d]^k$, and satisfies $\bT_\uu = \bT_{\uu'}$ whenever $\uu'$ is a permutation of $\uu$. Section~\ref{sec:Unfold.Tensor} treats general $k\ge3$ under rank and incoherence assumptions. Section~\ref{sec:Overcomplete} treats a model of random tensors for $k=3$.

\subsection{Notions of tensor rank}

As mentioned in the introduction, there are a few common non-equivalent ways to formalize the notion of rank for a (non-zero) tensor $\bT\in\R^{d_1}\otimes \cdots\otimes \R^{d_k}$.
In this paper, we define the \emph{rank} of $\bT$ as the minimum integer $r$ such that
$\bT$ can be expressed as a sum of $r$ pure tensors:
	\[r(\bT) \equiv 
	\min\Big\{m\ge1 :
	\bT = \sum_{s=1}^m v_s^{(1)}
		\otimes\dots\otimes v_s^{(k)}
	\textup{ for }
	v^{(\ell)}_i\in\R^{d_i}\Big\}\,.\]
We omit the argument $\bT$ whenever it is clear from the context.

A different notion of rank, which is also common in the literature, is given by considering --- for each $1\le i\le k$ ---
the matrix
$X^{(i)}\equiv\unfold^{(i)}(\bT)$
of dimensions
$d_i \times ((d_1\cdots d_k)/d_i)$,
with entries
	\[
	(X^{(i)})_{u_i,\uu_{-i}}
	=\bT_{\uu}
	\]
where $\uu_{-i}$
is $\uu$ without its $i$-th index. Write $\spn^{(i)}(\bT)$ for the column space of $X^{(i)}$, and define
	\beq\label{e:tucker}
	\rmu{i}(\bT)
	\equiv\dim\spn^{(i)}(\bT)
	=\rank X^{(i)}\,.
	\eeq
The \emph{multilinear rank}
or \emph{Tucker rank} of $\bT$
is defined as $\rmum(\bT)
=\max\set{\rmu{i}(\bT) : 1\le i\le k}$.
Again, we omit the argument $\bT$ whenever it is clear from the context. It is clear from the definition that $\rmu{i}\le \max\set{r,d_i}$. On the other hand we have
	\beq\label{e:r.ubd.tucker.rank}
	r\le (\rmu{1} \cdots \rmu{k})/(\rmum)
	\le (\rmum)^{k-1}\,;
	\eeq
we prove this fact in the appendix  (Lemma~\ref{l:tucker}).

\section{Tensor completion via unfolding}
\label{sec:Unfold.Tensor}

In this section we assume a symmetric underlying tensor $\bT\in (\R^d)^{\otimes k}$,
with $k\ge3$. We observe a subset of entries $\ee\subseteq [d]^k$ of size $\nn=|\ee|$, and denote
 the partially observed tensor
$\bY = \Proj_\ee(\bT)$. We now describe our algorithm, discuss our assumptions, and state 
our performance guarantees.
Proofs are in Appendix~\ref{appx:tensor.unfold}.

\subsection{Completion algorithm}
Our algorithm takes as input the set of indices $\ee$ and the partially observed tensor 
$\bY = \Proj_\ee(\bT)$. It also takes a threshold value $\lmstar$, which can be interpreted as a regularization parameter. 
In Theorem~\ref{t:tensor} we provide an explicit prescription for $\lmstar$ (see~\eqref{e:lmstar}).

\begin{alg}\label{alg.gen}
\begin{framed}
\noindent\textbf{Tensor completion via unfolding.} Input: $\ee$, $\bY$, $\lmstar$.
\begin{enumerate}[1.]
\item \emph{Sample splitting.} Partition the observed entries $\ee$ in two disjoint subsets
$\ee_1,\ee_2$ uniformly at random,
subject to $|\ee_1|=|\ee_2|=\nn/2$. 
Let $\delta_1\equiv \nn/(2d^k)$.
Denote by $\bY_1 = \Proj_{\ee_1}(\bY)$, $\bY_2 = \Proj_{\ee_2}(\bY)$
the corresponding partially observed tensors.

\item \emph{Tensor unfolding.} Set $a= \lfloor k/2\rfloor$,  $b=\lceil k/2\rceil$, and let $Z = \unfold^{a\times b}(\bY_1)$. Use $Z$ to define
	\beq\label{eq:Bdef}
	B =
	\f{1}{\delta_1}\Proj(ZZ^\sT)
	+\f{1}{(\delta_1)^2}\Proj_\perp(ZZ^\sT)
	\,.
	\eeq

\item \emph{Spectral analysis.} Compute the eigenvectors of $B$ with eigenvalues $\ge\lmstar$, and let $Q:\R^{d^a}\to\R^{d^a}$ be the orthogonal projection onto their span.

\item \emph{Denoising.} Let $\cQ:(\R^d)^{\otimes k}\to (\R^d)^{\otimes k}$
be the orthogonal projection defined by
\beq\label{e:tensor.right.Q}
\cQ =
\left\{\hspace{-4pt}\begin{array}{ll}
	Q\otimes Q\otimes Q & \text{if }k=3,\\
	Q\otimes Q & \text{if $k\ge 4$ even,}\\
       Q\otimes Q\otimes \id_{d} & \text{if $k\ge 5$ odd.}
\end{array}\right.\eeq
Let $\delta_2=\delta_1/(1-\delta_1)$, 
and let
$\hbT\equiv \bY_1
+(\delta_2)^{-1}\bY_2$.
Return the tensor
$\hbTstar = \cQ(\hbT)$.
\end{enumerate}
\end{framed}
\end{alg}

As we already commented, our algorithm differs from standard unfolding-based methods in that it does not seek to directly complete the tensor matricization, but only to estimate its left singular space. Completion is done by a ``denoising'' procedure which uses this singular space estimate, but also takes advantage of the original tensor structure.

\subsection{Rank and incoherence assumptions}

We will analyze the performance of Algorithm~\ref{alg.gen}  subject to rank and incoherence conditions which we now describe. In particular, we allow for a slightly less restrictive notion of rank.

\begin{ass}\label{ass:Main} We say that a tensor $\bT\in(\R^d)^{\otimes k}$ has \emph{unfolding parameters} $(\RU,\ALPHA,\MU)$ if, for $a=\lfloor k/2\rfloor$ and $b=\lceil  k/2\rceil$, the matrix $X = \unfold^{a\times b}(\bT)$ satisfies
\begin{enumerate}
\renewcommand{\labelenumi}{\textbf{(\theenumi)}}
\renewcommand{\theenumi}{\textup{\textbf{T\arabic{enumi}}}}
\item\label{e:assump.rank}
	$\rank X\le\RU$;
\item\label{e:assump.max.frob} 
	$d^k \|X\|_\infty^2 \le \ALPHA \|X\|_\F^2$;
\item\label{e:assump.frob.op}
	$\MU\|X\|_\F^2=\RU\|X\|_\op^2$.
\end{enumerate}
Note that \ref{e:assump.rank}~and~\ref{e:assump.max.frob} 
are inequalities but \ref{e:assump.frob.op} is an equality.
\end{ass}

\begin{rmk}\label{r:WLOG.coherence.tensor}
A few comments are in order. First of all,
note that $\rmum(\bT)\le \RU(\bT)\le r(\bT)$, which means that \ref{e:assump.rank} is less restrictive than the assumption $r(\bT)\le\RU$. Next, since $\|X\|_\infty^2 \le \|X\|_\F^2 \le d^k \|X\|_\infty^2$, we can assume
$1\le \ALPHA\le d^k$; it is standard in the literature to assume that $\ALPHA$ is not too large. Lastly, since $\|X\|_\op^2\le\|X\|_\F^2\le\RU\|X\|_\op^2$, we can assume $1\le\MU\le\RU$.
\end{rmk}

With these definitions, we can now state our result
on the guarantees of Algorithm~\ref{alg.gen}. Define
	\beq\label{e:r.max}
	r_{\max}(d;k)
	= \left\{\begin{array}{ll}
	d^{3/4} & k=3,\\
	d^{k/2} & k\ge 4 \textup{ even},\\
	d^{k/2-1} & k\ge 5 \textup{ odd},\\
	\end{array}\right.
	\eeq

\begin{thm}\label{t:tensor} 
Let $\bT\in (\R^d)^{\otimes k}$ be a deterministic symmetric tensor satisfying Assumption~\ref{ass:Main} with unfolding parameters $(\RU,\ALPHA,\MU)$, such that $\RU\le r_{\max}$. Suppose that we observe $\nn$ entries of $\bT$
uniformly at random. Let $\hbTstar$ be the spectral estimator of  Algorithm~\ref{alg.gen} with
	\beq\label{e:lmstar}
	\lmstar=4(k\log d)^8
	\bigg( \f{\ALPHA\RU \MU^{1/2}}{\nn/d^{k/2}}
	\bigg)^{ 2/3}
	\|B\|_\op\,.\eeq
Then, in the regime $32(k\log d)^{12}\ALPHA\RU \MU^{1/2} d^{k/2}\le \nn\le (k\log d)^{16} \ALPHA\RU\MU^2 d^b$, we have
	\beq\label{eq:MainBound}
	\|\hbTstar-\bT\|_\F\le
	20 (k\log d)^{3}
	\bigg(\frac{\ALPHA\RU \MU^2}
		{\nn /d^{k/2} }\bigg)^{1/3}
	\|\bT\|_\F\,.
	\eeq
with very high probability.
\end{thm}

Theorem~\ref{t:tensor} shows that
a symmetric
rank-$r$ tensor $\bT\in(\R^d)^{\otimes k}$
can be reconstructed
by spectral methods,
based on $\nn\gtrsim r d^{k/2}$ revealed entries. Apart from logarithmic factors, we suspect that this condition on $\nn$ 
may be optimal among polynomial-time methods.
One supporting evidence is that, for $k=3$,
this matches the bounds
\eqref{e:bm.main.lbd} and \eqref{e:bm.random.lbd}
of the degree-$6$ \textsc{sos} algorithm \cite{BarakMoitra}.
The authors further prove
(\cite[Theorem 3]{BarakMoitra})
that their condition~\eqref{e:bm.main.lbd}
under Feige's hypothesis
\cite{feige} on the refutation of random satisfiability formulas. On the other hand, the error bound
\eqref{eq:MainBound} is quite possibly suboptimal, arising as an artifact of the algorithm or of the analysis.
We believe that our rank condition $\RU\le r_{\max}$ is also suboptimal;
for algorithms of this type
the tight condition seems likely to be
of the form
$\RU\le d^{\lfloor k/2\rfloor}$
(maximum rank of the unfolding).

\section{Overcomplete random three-tensors}
\label{sec:Overcomplete}

In this section we describe our algorithm for overcomplete three-tensors, and state its guarantees 
for a certain random tensor model.
Proofs are in Appendix~\ref{appx:three}.

\subsection{Completion algorithm}

Algorithm~\ref{alg.gen} of Section~\ref{sec:Unfold.Tensor} is limited to tensors $\bT$ with rank $r\ll r_{\max}$, as defined in \eqref{e:r.max}.  As we already noted above, this particular condition is most likely suboptimal.
However, among all algorithms of this type
(i.e., based on spectral analysis of the unfolded tensor), we expect that a fundamental barrier is $r \ll d^{\lfloor k/2\rfloor}$.
Beyond this point, the unfolded tensor has nearly full rank, and we do not expect the projector $Q$ to have helpful denoising properties.

On the other hand, the number of parameters required to specify a rank-$r$ tensor in $(\R^d)^{\otimes k}$ is of order $rd$, so we might plausibly hope to complete it given $\nn \gg rd$ entries. This only imposes the rank bound $r \ll d^{k-1}$. In this section we consider the case $k=3$: from the above argument the information-theoretic bound is $r \ll d^2$. Our unfolding method (Algorithm~\ref{alg.gen}) can complete the tensor  up to rank $r\ll d^{3/4}$, by Theorem~\ref{t:tensor}. From the preceding discussion, this bound is likely to be  suboptimal, but the best we expect from such an algorithm is $r\ll d^{\lfloor 3/2\rfloor}=d$.

Motivated by these gaps, in this section we
develop a different completion algorithm for the case $k=3$, which avoids unfolding and relies instead on a certain ``contraction'' of the tensor with itself. This was motivated by ideas developed in~\cite{hsss} for the tensor decomposition problem.
Under a natural model of random
symmetric low-rank tensors, we prove
that in the regime $d\le r\ll d^{3/2}$,
our algorithm succeeds in completing the tensor based on $\nn\gg rd^{3/2}$ observed entries.

The algorithm takes as input the set of observed indices $\ee$,
the partially observed tensor
$\bY = \Proj_\ee\bT$,
and a threshold value $\lmstar$.
In Theorem~\ref{t:three} we provide an explicit prescription for $\lmstar$.

\begin{alg}\label{alg.three}
\begin{framed}
\textbf{Completion for three-tensors
	via contraction.} 
Input: $\ee,\bY,\lmstar$.
\begin{enumerate}[1.]
\item \emph{Sample splitting.} 
Let $\delta$ be defined by the relation
$1-(1-\delta)^3 = |\ee|/d^3$. Take subsets $\II,\JJ,\KK\subset\ee$ which are uniformly random subject to the following conditions:
each $\II$, $\JJ$, $\KK$ has size $d^3\delta$;
each pairwise intersection
$\II\cap\JJ$, $\II\cap\KK$, $\JJ\cap\KK$
has size $d^3\delta^2$;
the triple intersection
$\II\cap\JJ\cap\KK$ has size $d^3\delta^3$.
(This implies, in particular, that $\II\cup\JJ\cup\KK=\ee$.)
Denote the corresponding partially observed tensors
$\dot{\bY}=\Proj_\II\bT$,
$\ddot{\bY}=\Proj_\JJ\bT$,
and $\dddot{\bY}=\Proj_\KK\bT$.

\item \emph{Tensor contraction.} Let
$W$ be the $d^2\times d^2$ matrix with entries
	\beq\label{e:def.contract.tensor}
	W_{\ui,\uj}
	= \f1{\delta^2}
	\sum_{\ell\le d}
	\dot{\bY}_{\ell i_1 j_1}
	\ddot{\bY}_{\ell i_2 j_2}\,.
	\eeq

\item \emph{Spectral analysis.} 
Compute the singular value decomposition of $W$.
Take the singular vectors of $W$ with singular values $\ge\lmstar$, and let 
$Q:\R^{d^2}\to\R^{d^2}$ 
be the orthogonal projection onto their span.

\item \emph{Denoising.} Let $\cQ :(\R^d)^{\otimes 3}\to (\R^d)^{\otimes 3}$ be defined by $\cQ = Q\otimes \id_d$. Let $\hbT = \delta^{-1}\dddot{\bY}$, and return the tensor $\hbTstar = \cQ(\hbT)$.
\end{enumerate}
\end{framed}
\end{alg}

\subsection{Random tensor model}

We analyze Algorithm~\ref{alg.three} in 
a random model:

\begin{ass}\label{ass:Random}
We say that $\bT\in (\R^d)^{\otimes 3}$ is a \emph{standard random tensor} with $r$ components if 
	 \beq\label{e:random.tensor}
	 \bT = \sum_{s\le r}
	 	a_s\otimes a_s\otimes a_s
	 \eeq
where $a_1,\ldots,a_r$ are i.i.d.\  random vectors in $\R^d$
such that $x=a_i$ satisfies the following: \begin{enumerate}
\renewcommand{\labelenumi}{\textbf{(\theenumi)}}
\renewcommand{\theenumi}{\textup{\textbf{A\arabic{enumi}}}}
\item\label{e:vector.symmetric}
	(symmetric)
	$x$ is equidistributed as $-x$;
\item\label{e:vector.isometric}
	(isometric)
	$\E[ x x^\sT ]=\id_d /d$;
\item\label{e:vector.subgaussian}
	(subgaussian)
	$\E[\exp(\<x,v\>)]
	\le \exp\{ \tau^2 \|v\|^2/(2d) \}$
	for all $v\in\R^d$.
\end{enumerate}
\end{ass}

Note Assumption~\ref{ass:Random} has a slight abuse of notation, since the tensor \eqref{e:random.tensor} can have, in general, rank smaller than $r$. However, in the regime of interest, we expect the rank of $\bT$ to be close to $r$ with  high probability.

\begin{thm}\label{t:three} Let $\bT\in(\R^d)^{\otimes 3}$  be a standard random tensor \eqref{e:random.tensor} satisfying Assumption \ref{ass:Random}. Suppose that we observe $\nn$  entries of $\bT$ uniformly at random, where $\nn\ge\max\set{r,d} d^{3/2}$ and $r\le d^2$. Let $\hbTstar$ be the spectral estimator of Algorithm~\ref{alg.three} with 
	\beq
	\label{e:three.lambda.star.in.n}
	\lmstar
	=\bigg(
	\f{ d^{3/2}\max\set{d,r}}
		{  \nn}
	\bigg)^{4/5}\,.
	\eeq
Then, with very high probability,
	\beq
	\label{e:three.error.bound}
	\|\hbTstar-\bT\|_{\F}
	\lesssim
	\bigg(
	\f{d^{3/2}\max\set{d,r}}{\nn}
	\bigg)^{1/5}
	\|\bT\|_\F\,.
	\eeq
\end{thm}

If one uses crude matrix calculations (not taking advantage of the sparsity or low rank of the matrices involved), we estimate the runtimes of our methods as follows. In Algorithm~\ref{alg.gen}, computing the matrix $B$ of \eqref{eq:Bdef} takes time $O(d^{k+b})$; finding its eigendecomposition takes time $O(d^{3a})$; and the denoising step can be done in time $O(d^{k+a})$. Thus the overall runtime is $O(d^{k+b})$, which for $k=3$ becomes $O(d^5)$. In Algorithm~\ref{alg.three}, computing the matrix $W$ of \eqref{e:def.contract.tensor} takes time $O(d^5)$; finding its singular value decomposition takes time $O(d^6)$; and the denoising step can be done in time $O(d^4)$. Thus the overall runtime is $O(d^6)$; so Algorithm~\ref{alg.gen} is preferable when the rank is low.

\section{Numerical illustration}
\label{sec:Numerical}
\def\mse{\textup{\textsf{\footnotesize MSE}}}

We illustrate our results with numerical simulations of random tensors
	\beq\label{e:random.gaus.tensors}
	\bT = \sum_{s\le r}
	a_s\otimes a_s\otimes a_s\,.\eeq
We assume (cf.\ Assumption~\ref{ass:Random}) that $a_1,\ldots,a_r$ are independent  gaussian random vectors in $\R^d$, with $\E a_s=0$ and $\E(a_s(a_s)^\sT)=\id_d/d$. Our simulations estimate the normalized mean squared error
	\beq\mse \equiv 
	\f{\E (\|\hbTstar-\bT\|_{\F}^2 )}
	{\E(\|\bT\|_{\F}^2)} \, ,
	\label{eq:MSEdef}\eeq
where $\hbTstar$ is the output of the completion algorithm.
	
\subsection{Performance of unfolding algorithm}

\begin{figure}[h!]
\begin{subfigure}[h!]{.49\textwidth}
\centering
\includegraphics[width=\textwidth]{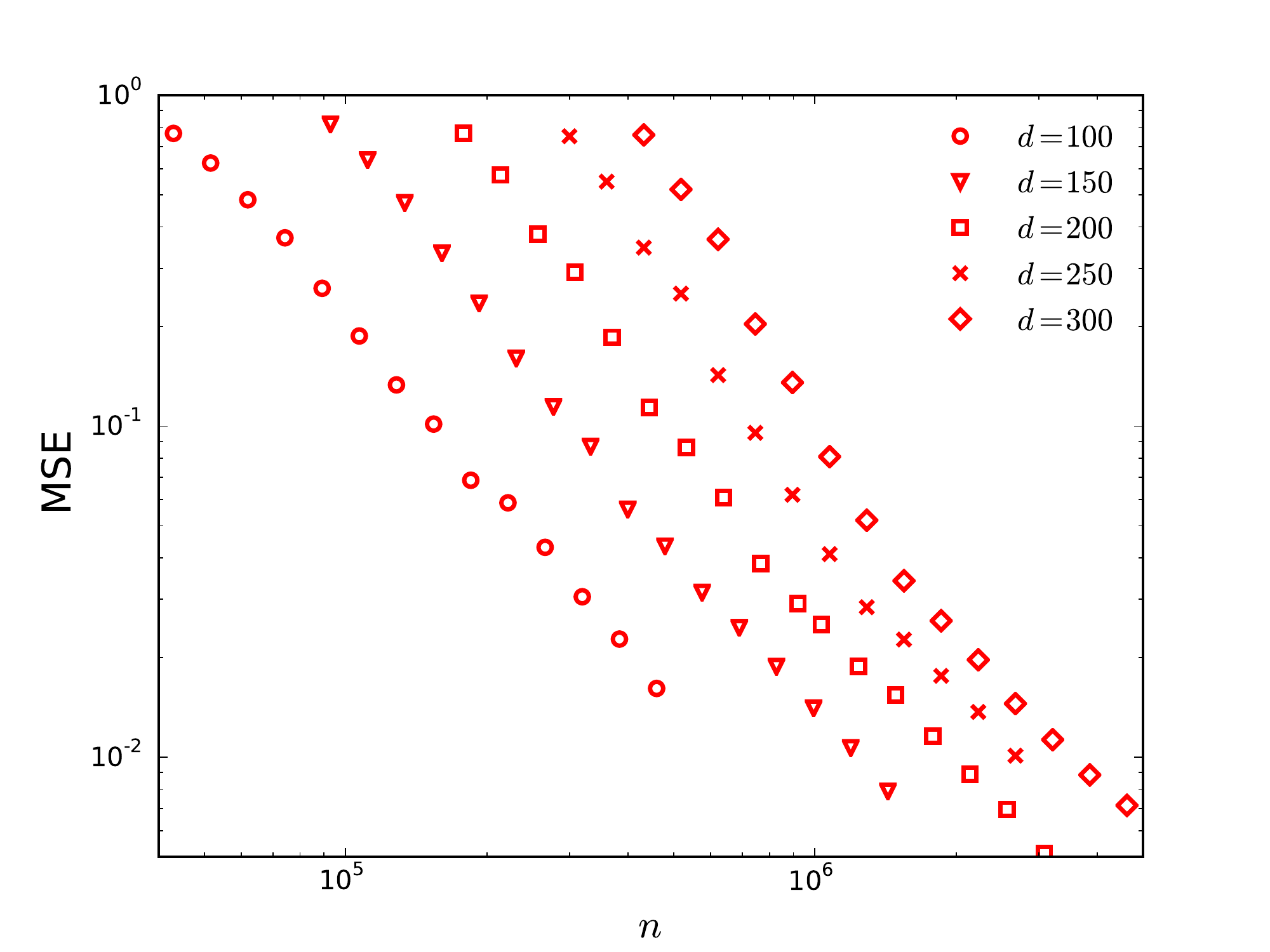}
\caption{Estimated $\mse$ vs.\
sample size $\nn$.}
\end{subfigure}
\begin{subfigure}[h!]{.49\textwidth}
\centering
\includegraphics[width=\textwidth]{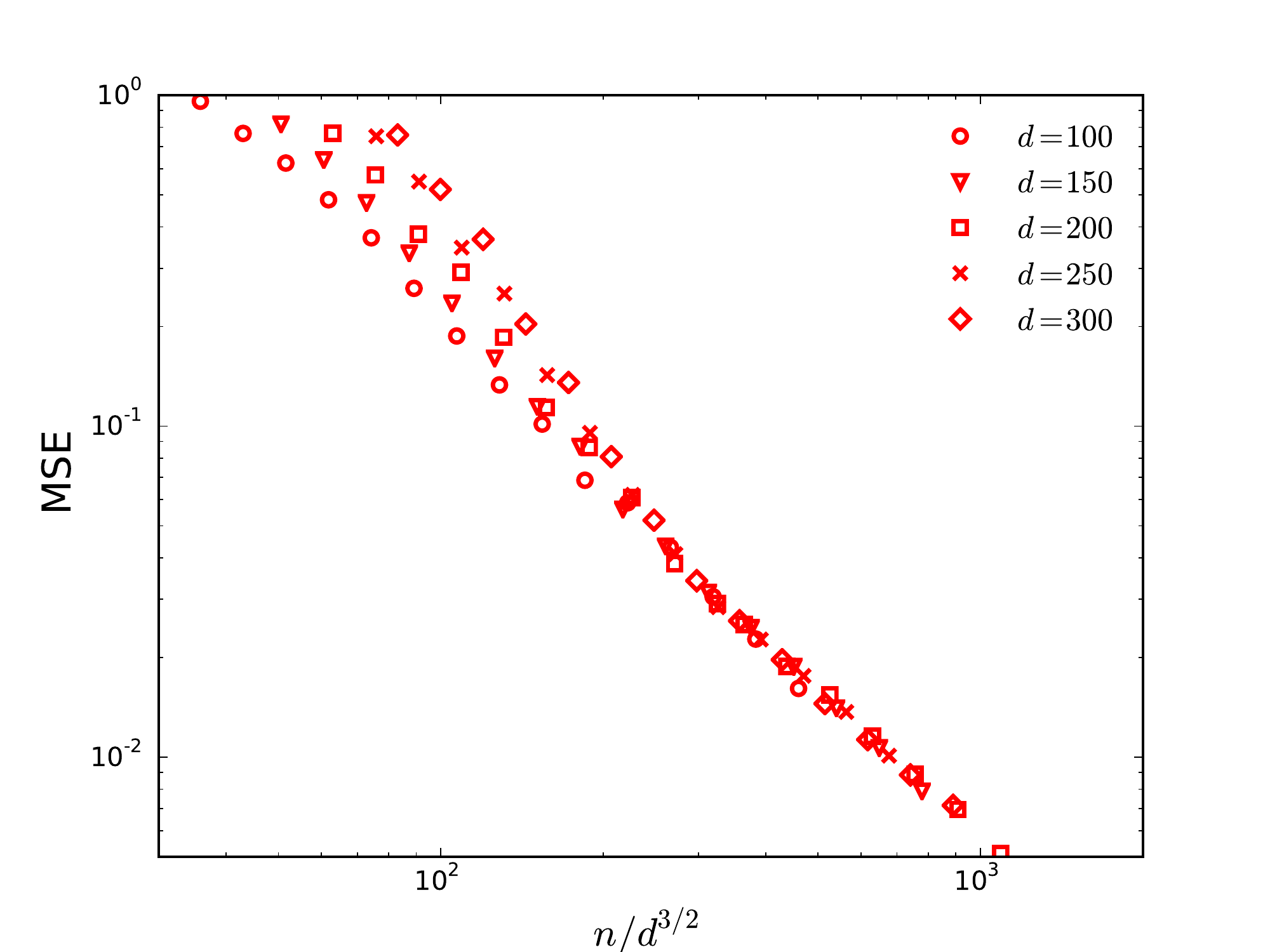}
\caption{Estimated $\mse$ vs.\ rescaled
sample size $\nn/d^{3/2}$.}
\end{subfigure}
\caption{Numerical illustration of Algorithm~\ref{alg.gen} for completing random tensors \eqref{e:random.gaus.tensors} of 
order $k=3$ and rank $r=4$. Performance is measured in terms of normalized
mean squared error $\mse$ (see~\eqref{eq:MSEdef}),
estimated by averaging over $100$ realizations.}\label{fig:TensorSimple}
\end{figure}

Figure~\ref{fig:TensorSimple} reports the performance of our unfolding method (Algorithm~\ref{alg.gen}) in the undercomplete regime, taking $r=4$. We plot the normalized mean square error \eqref{eq:MSEdef} estimated by averaging over $100$ independent random realizations of $\bT$ and of the set $\ee$ of revealed entries. We set the threshold parameter
	\beq\label{eq:Lstar}
	\lambda_\star = 
	3 ( d^{3/2}/\nn)^{2/3} \|B\|_{\op}
	\eeq
--- this choice was guided by the prescription~\eqref{e:lmstar} of Theorem~\ref{t:tensor}, as follows:
in the present setting, we have $X
= \unfold^{1\times 2}(\bT)$. If we write $f\sim g$ to indicate $\lim_{d\to\infty} f(d)/g(d)=1$, then
	\begin{align*}
	\|X\|_{\op} &= \max_{i\le r}
	\|a_i(a_i\otimes a_i)^{\sT}\|_{\op}
	\sim1\, , \\
	\|X\|_{\F} & =
	\sum_{s\le r}\sum_{t\le r}
		\<a_s,a_t\>^3
		= \sum_{s\le r}\|a_s\|_2^
		6 +O(r^2/d^{3/2}) 
		\sim r\, ,\\
	\|X\|_{\infty}
	&=
	\max_{i,j,l\le d}
	\bigg|
	\sum_{s=1}^r a_{s,i}a_{s,j} a_{s,l}
	\bigg|
	\sim\max_{s\le r,i\le d} |a_{s,i}|^3
	\sim (2\log d)^{3/2}\,.
	\end{align*}
Therefore $X$ satisfies Assumption~\ref{ass:Main}
with $\RU=r$, $\MU\sim1$, and $\ALPHA
\sim (2\log d)^3/r$. Our choice of the parameter $\lmstar$
is obtained by substituting these into \eqref{e:lmstar}. After some trial and error, we chose the factor $3$ in \eqref{eq:Lstar} instead of logarithmic factors, which appeared to be overly pessimistic for moderate values of $d$.

\subsection{Performance of spectral algorithm for overcomplete tensors}  
\begin{figure}[h!]
\begin{subfigure}[h!]{.49\textwidth}
\centering
\includegraphics[width=\textwidth]{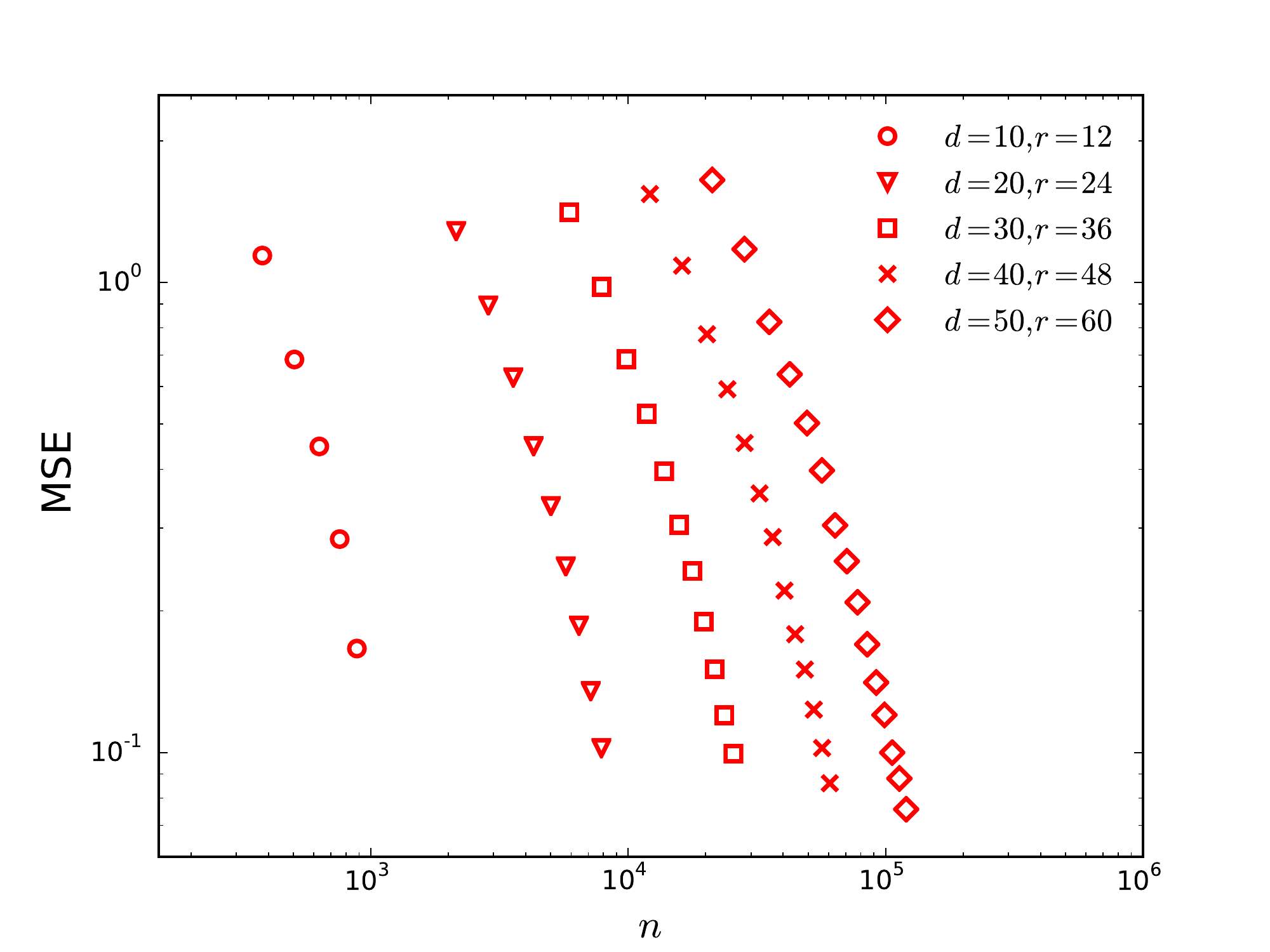}
\caption{Estimated $\mse$ vs.\ sample size $\nn$.}
\end{subfigure}
\begin{subfigure}[h!]{.49\textwidth}
\centering
\includegraphics[width=\textwidth]{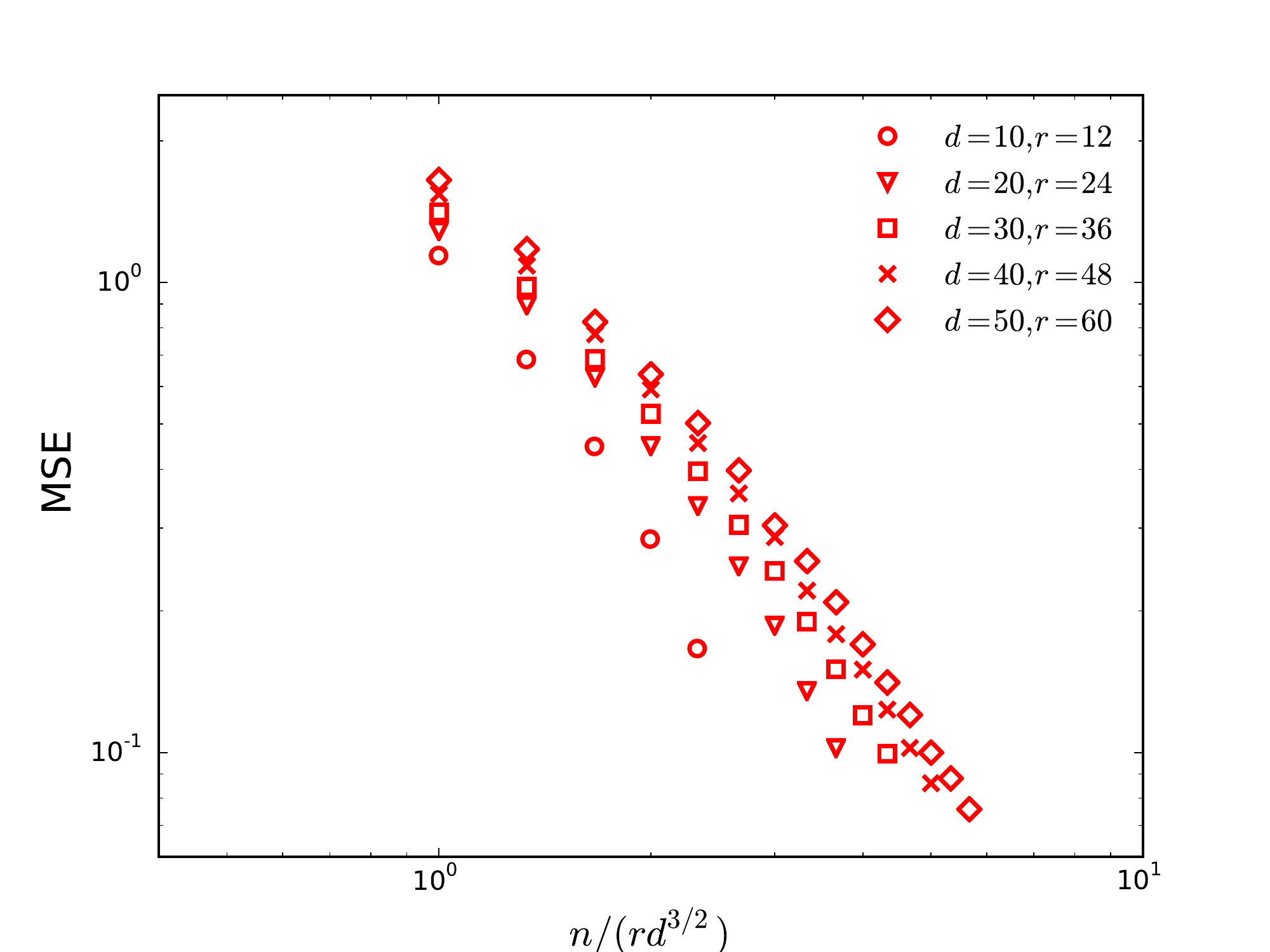}
\caption{Estimated $\mse$ vs.\ 
rescaled sample size $\nn/d^{3/2}$.}
\end{subfigure}
\caption{Performance of Algorithm 2 in completing \emph{overcomplete} random tensors of rank $4$, and order $k=3$. Left frame:
mean square error of reconstruction, estimated by averaging over $100$ realizations, plotted against the number of revealed entries. 
Right frame: same data plotted against the rescaled number of revealed entries $\nn/(rd^{3/2})$.
}\label{fig:TensorOver}
\end{figure}

Figure~\ref{fig:TensorOver} reports the performance of our spectral method for the overcomplete regime (Algorithm~\ref{alg.three}), taking $r/d=1.2$. We set $\lmstar$ according to the prescription~\eqref{e:three.lambda.star.in.n} of Theorem~\ref{t:three}. For each value of $d$, the $\mse$ appears to decrease rapidly with $\nn$. The plots (for various values of $d$) of the $\mse$ versus the rescaled sample size $\nn/(r d^{3/2})$ appears to approach a limiting curve.  This suggests that the threshold for our method to succeed in reconstruction occurs around $\nn=rd^{3/2}$, which is consistent with the bound of our Theorem~\ref{t:three}.

\section{Column spaces
of partially revealed wide matrices}
\label{sec:Matrix}

In this section we present our results on the column spaces of partially revealed $d_1\times d_2$ matrices. As mentioned above, these results are the main input to the proof of Theorem~\ref{t:tensor}. The conclusions obtained in this section are most interesting for the regime $d_1 \ll d_2$. 

\subsection{Incoherence condition}

\begin{ass}\label{d:incoherent}
We say that a matrix $X\in\R^{d_1\times d_2}$ is
\emph{$(\LL,\MM,\RR)$-incoherent} if
\begin{enumerate}
\renewcommand{\labelenumi}{\textbf{(\theenumi)}}
\renewcommand{\theenumi}{\textbf{M\arabic{enumi}}}
\item \label{e:incoherent.rows}
$d_1 \cdot\max_i \|X^\sT e_i\|^2
		\le\LL\|X\|_\op^2$;
\item \label{e:incoherent.cols}
	$d_2\cdot \max_j\|Xe_j\|^2
	\le\RR\|X\|_\op^2$; and
\item \label{e:incoherent.max}
	$d_1d_2 \cdot\max_{i,j} |X_{i,j}|^2
	\le\LL\MM\RR\|X\|_\op^2$,
\end{enumerate}
where $1\le i\le d_1$ and $1\le j\le d_2$.
\end{ass}

It is easily seen (cf.\ Lemma \ref{lemma:WLOG_Coh}) that one can assume without loss of generality $1/d_1\le\LL\le d_1$,  
$1/d_2\le\RR\le d_2$,  $1\le \LL\MM\RR \le d_1d_2$. To motivate the above condition, we observe that
it can be deduced as a consequence of a standard incoherence assumption, that we recall below.

\begin{dfn}[\cite{MR2565240}]
\label{d:coherence}
Let $W$ be an $r$-dimensional subspace of 
$\R^d$, and let $P_W$ be the orthogonal projection onto $W$. The \emph{coherence} of $W$ (with respect to the standard basis $(e_i)_{i\le d}$ of $\R^d$) is
	\[\coher_W
	= \f{d}{r}\max_{1\le i\le d} 
	\|P_W e_i\|^2\,.\]
Note the trivial bounds
	\[
	\f{d}{r} 
	\ge \coher_W
	\ge \f{1}{r} \sum_{i=1}^d\|P_We_i\|^2
	=1\,.
	\]
If $M$ is a $d\times r$ matrix whose columns form an orthonormal basis of $W$, then we can express $P_W=MM^\sT$, so that
	\[\coher_W
	=\f{d}{r}\max_{1\le i\le d} 
	\|MM^\sT e_i\|^2
	=\f{d}{r}\max_{1\le i\le d} 
	\sum_{s=1}^r
	(M_{is})^2\,.\]
We denote $\coher_M\equiv \coher_W$.
\end{dfn}

We refer to \cite{MR2565240} for further discussion of this coherence condition, which has become fairly standard in the literature. We now give two illustrations for 
Assumption~\ref{d:incoherent}:

\begin{enumerate}[1.]
\item \emph{Derivation of Assumption~\ref{d:incoherent} from
	Definition~\ref{d:coherence}.}\\
Suppose $X$ is $d_2\times d_2$ with singular value decomposition $UDV^\sT$, where $U^\sT U = \id_\RU= V^\sT V$ and $D$ is diagonal. One can then easily verify that $X$ is $(\LL,\MM,\RR)$-incoherent with 
	\beq\label{e:coherence.params.reln}
	\LL=(\coher_U)\RU,\quad
	\RR=(\coher_V)\RU,\quad
	\MM=1\,.
	\eeq
(see Lemma~\ref{lemma:IncoherenceComparison} for the proof). That is to say, imposing Assumption~\ref{d:incoherent} with $\LL=\RR=c \RU$ and $\MM=1$ is less restrictive than imposing that $X$ is of rank $\RU$ with $c$-incoherent singular vectors.

\item \emph{Derivation of Assumption~\ref{d:incoherent} from
Assumption~\ref{ass:Main}.}\\ Alternatively, suppose $X$ satisfies an entrywise bound
$d_1 d_2\|X\|_\infty^2 \le\bar{\varpi}\|X\|_\op^2$.
It is then trivial to verify that $X$ is $(\LL,\MM,\RR)$-incoherent with 
	\beq\label{e:coherence.params.reln.if.bdd}
	\LL=\RR=1/\MM = \bar{\varpi}.
	\eeq
In Assumption~\ref{ass:Main},  conditions~\ref{e:assump.max.frob}~and~\ref{e:assump.frob.op}
together imply (with $d_1=d^a$ and $d_2=d^b$)
	\[
	d_1d_2\|X\|_\infty^2
	\le\ALPHA\|X\|_\F^2
	\le (\ALPHA\RU/\MU) \|X\|_\op^2\,,
	\]
so we have
\eqref{e:coherence.params.reln.if.bdd} with $\bar{\varpi}=\ALPHA\RU/\MU$. That is to say, imposing Assumption~\ref{d:incoherent} with $\LL=\RR=1/\MM=\ALPHA\RU/\MU$ is less restrictive than imposing Assumption~\ref{ass:Main} with parameters $(\RU,\ALPHA,\MU)$.
\end{enumerate}
In the tensor completion problem we work with the second scenario \eqref{e:coherence.params.reln.if.bdd}.

\subsection{Estimation error}

We now state our main result on column space estimation for partially revealed matrices.

\begin{thm}\label{t:op.error} Suppose that $X\in \R^{d_1\times d_2}$ is $(\LL,\MM,\RR)$-incoherent. Let $\ee\subseteq[d_1]\times[d_2]$ be the random set of observed entries, where each $(i,j)\in[d_1]\times[d_2]$ is included in $\ee$ independently with probability $\delta= \nn/(d_1d_2)$. Given the observed matrix $Y\equiv\Proj_\ee X$, let
	\beq\label{e:def.hB}
	\hB\equiv
	\f1\delta \Proj(YY^\sT)
	+\f1{\delta^2} \Proj_\perp(YY^\sT)\,.
	\eeq
Then, for $d_1d_2\ge 3000$, we have
	\beq\label{e:BoundMatrix}
	\f{\|\hB-XX^\sT\|_\op}
	{ 2(\log(d_1d_2))^4 }
	\le 
	\f{(\LL \RR d_1 d_2)^{1/2}
		\|X\|_\op^2}{\nn}
	\max\bigg\{
	1\,, \bigg(\f{\nn}{d_2\LL}\bigg)^{1/2}\,,
	\f{(\LL\MM^2\RR d_1 d_2)^{1/2}}{\nn}\,,
	\bigg(\f{\LL\RR}{d_2}\bigg)^{1/2}
	\bigg\}
	\eeq
with probability at least
$1-4d_1\exp\{-\tfrac18\log(d_1d_2)^2\}$.
\end{thm}

\begin{cor}\label{c:op.error.simp} In the setting of Theorem~\ref{t:op.error}, assume additionally that
$(\LL\MM^2\RR d_1d_2/t)^{1/2} \le \nn \le t \LL d_2$
and 
$\MM\RR\le t d_2$. Then the error bound \eqref{e:BoundMatrix} simplifies to
	\[
	\f{\|\hB-XX^\sT\|_\op}
	{2(\log(d_1d_2))^4}
	\le \f{(t \LL\RR d_1 d_2)^{1/2}}{\nn}
	\|X\|_\op^2\,.\]

\begin{proof}
Immediate consequence of Theorem~\ref{t:op.error}.
\end{proof}
\end{cor}

From our perspective, the most interesting application of the above is as follows.
Recalling \eqref{e:coherence.params.reln},
suppose the $d_1\times d_2$ matrix $X$
is $(\LL,\MM,\RR)$-incoherent with $\LL=\RR=cr$ 
and $\MM=1$. Consider Corollary~\ref{c:op.error.simp}
with $t=1$: then the conditions reduce to
 $cr \le (d_2)^{1/2}$
and $cr (d_1 d_2)^{1/2} \le \nn \le crd_2$,
where the latter can only be satisfied
if $d_1 \le d_2$. With these conditions,
Corollary~\ref{c:op.error.simp} says that the column space of $X$ can be well approximated by the top eigenvectors of the matrix $\hB$, provided we saw (roughly) $\nn \gg r(d_1 d_2)^{1/2}$ entries.
We emphasize that this result implies, for $d_1\ll d_2$, a wide regime of sample sizes
	\[r(d_1 d_2)^{1/2} \ll \nn \ll r d_2\]
from which we can obtain a good estimate of the sample space, even though it is impossible to complete the matrix (in the sense of Frobenius norm approximation). In this regime, the column space estimate can be useful for (partial) matrix completion: if $Q$ approximates projection onto the left column space of $X$, and $y$ is a column of $\Pi_\ee X$ containing $d\delta'\gg r$ observed entries, then $Qy/\delta'$ is a good estimate of the corresponding column of $X$.

\subsection*{Acknowledgements}

This research was partially supported by
NSF grant CCF-1319979 (A.M.)
and NSF MSPRF grant DMS-1401123 (N.S.).

{\raggedright
\bibliographystyle{alphaabbr}
\bibliography{refs}
}

\pagebreak

\appendix

\section{Standard matrix inequalities}
\label{s:std.matrix.ineq}

In this appendix we collect a few standard tools that will be used several times in our proofs.  For any real-valued random variable $X$, the \emph{essential supremum} $\esssup X$ is the minimal value $R$ such that $\P(X\le R)=1$. 
Recall the following form of the Chernoff bound: if $X$ is a binomial random variable with mean $\bar{\mu}$, then for all $t\ge1$ we have
	\beq\label{e:chernoff}
	\P(X\ge t\bar{\mu})
	\le \exp\{-t\bar{\mu}\log(t/e)\}\,.
	\eeq

\begin{ppn}[matrix Bernstein, rectangular  {\cite[Theorem~1.6]{MR2946459}}]\label{p:bernstein.rect}
Let $(Z_\ell)$ be a finite sequence of independent random
$d_1\times d_2$ matrices.
Assume $\E Z_\ell=0$ for all $\ell$, and let
	\beq\label{e:bernstein.rect.R.sigma}
	R =\max_\ell
	\bigg\{\esssup\|Z_\ell\|_\op\bigg\}\,,
	\quad
	\sigma^2
	=\max \bigg\{
	\bigg\|\sum_\ell \E[ Z_\ell(Z_\ell)^{\sT} ]
	\bigg\|_\op,
	\bigg\|\sum_\ell \E[ (Z_\ell)^{\sT}Z_\ell ]
	\bigg\|_\op\bigg\}\,.
	\eeq
Then, for all $t\ge0$,
	\[\P\bigg( \bigg\|
	\sum_\ell Z_\ell\bigg\|_\op\ge t \bigg)
	\le (d_1+d_2)
	\exp\bigg\{ -\f38
		\min\bigg\{ 
		\f{t^2}{\sigma^2},\f{t}{R}
		\bigg\} \bigg\}\,.\]

\end{ppn}

\newcommand{\ALL}[1]{\mathcal{#1}}
\newcommand{\ONE}[1]{#1}
\newcommand{\ZRO}[1]{#1_\circ}

\begin{ppn}[\cite{MR0309968}]
\label{p:wedin}
Suppose that $\ALL{A}$ and $\ALL{B}$ are positive semidefinite matrices,
with singular value decompositions
 	\begin{align*}\begin{array}{rll}
	\mathcal{A} =  A+A_\circ
	&\hspace{-6pt}=
	U\Sigma U^\sT
		+U_\circ\Sigma_\circ (U_\circ)^\sT,
	&
	\begin{pmatrix}
	U & U_\circ
	\end{pmatrix}^\sT
	\begin{pmatrix}
	U & U_\circ
	\end{pmatrix} = \id_r,\\
	\mathcal{B} = B + B_\circ
	&\hspace{-6pt}=
	O \Lambda O^\sT
	+O_\circ \Lambda_\circ (O_\circ)^\sT,
	&\begin{pmatrix}
	O & O_\circ
	\end{pmatrix}^\sT
	\begin{pmatrix}
	O & O_\circ
	\end{pmatrix} = \id_s.
	\end{array}
	\end{align*}
Suppose $\|\mathcal{A}-\mathcal{B}\|_\op\le\epsilon$,
and that
the maximum diagonal entry of $\Sigma_\circ$ is at most $\sigma$ while
the minimum diagonal entry of $\Lambda$
is at least $\sigma+\delta>0$. Then
	\[|\sin\theta(O,U)|
	\equiv\|(I-UU^\sT)OO^\sT\|_\op
	\le \epsilon/\delta\,.\]
\end{ppn}

\begin{ppn}[{\cite[Propn.~A.7]{hsss}}]
\label{p:trunc.bernstein} Let $(Z_\ell)_{\ell\le L}$ be a sequence of independent random
$d_1\times d_2$ matrices. Assume
 $\E Z_\ell=0$ for all $\ell$, and furthermore that
	\beq\label{e:bernstein.truncation}
	\P(\|Z_\ell\|_\op\ge\beta)\le p
	\quad\text{and}
	\quad
	\|\E[ Z_\ell
	\Ind{\|Z_\ell\|_\op\ge\beta} ]
	\|_\op
	\le q\,.
	\eeq
Denote $\sigma^2$ as in \eqref{e:bernstein.rect.R.sigma}. Then, for all $t\ge0$,
	\[\P\bigg(
	\bigg\|
	\sum_{\ell\le L} Z_\ell \bigg\|_\op
	\ge t+Lq
	\bigg)
	\le Lp
	+(d_1+d_2)
	\exp\bigg\{ -\f38
	\min\bigg\{
	\f{t^2}{\sigma^2},
	\f{t}{\beta}
	\bigg\}\bigg\}\,.\]
\end{ppn}

\begin{ppn}[matrix Rademacher, symmetric 
{\cite[Thm.~1.2]{MR2946459}}]
\label{p:rademacher} Let $(Z_\ell)$ be a finite sequence of $d\times d$ symmetric matrices. Let $(\sgns_\ell)$ be a sequence of independent symmetric random signs. Then
	\[\P\bigg(\bigg\|
	\sum_\ell\sgns_\ell Z_\ell
	\bigg\|_\op \ge t\bigg)
	\le 2d \exp\{-t^2/(2\sigma^2)\},
	\quad
	\sigma^2=\bigg\|\sum_\ell (Z_\ell)^2
	\bigg\|_\op\,.\]
\end{ppn}
\begin{ppn}[matrix decoupling 
{\cite[Thm.~1]{MR1334173}}
(see also {\cite[Thm.~5.13]{hsss}})]
\label{p:decoupling}
Let $(Z_{ij})$ be a family of matrices,
and let $(\sgns_i)$ and $(\sgnt_i)$ be
 sequences of  independent 
symmetric random signs. There is  an absolute
 constant $C$ such that for all $t\ge0$, 
	\[\P\bigg( \bigg\|
	\sum_{i\ne j}\sgns_i \sgns_j Z_{ij}\bigg\|_\op
	 \ge t\bigg)
	\le C \P\bigg( 
	\bigg\|\sum_{i\ne j}
	\sgns_i \sgnt_j
	Z_{ij}
	\bigg\|_\op
	\ge t\bigg)\,.\]
\end{ppn}

\section{Column space estimation
with large aspect ratios}
\label{s:matrix.completion}

In this appendix, we prove our matrix completion result, Theorem~\ref{t:op.error}. Before passing to the actual proof, we will establish some properties of the incoherence condition, Assumption \ref{d:incoherent}.

\subsection{Matrix incoherence conditions}

We begin by proving some easy observations
regarding our matrix incoherence conditions
(Assumption~\ref{d:incoherent}).

\begin{lem}\label{lemma:IncoherenceComparison}
Suppose $X\in\R^{d_1\times d_2}$ has singular value decomposition $UDV^\sT$, with $U^\sT U = \id_r = V^\sT V$. Then $X$ is $(\LL,\MM,\RR)$-incoherent with parameters $\LL=r \cdot \coher_U$, $\RR=r \cdot \coher_V$, and $\MM=1$.

\begin{proof} For indices $1\le i\le d_1$ and $1\le\ell\le d_2$ we have
	\begin{align*}
		|X_{i\ell}| 
		&= |\langle
		UU^\sT e_i, X VV^\sT e_\ell\rangle|
		\le \|X\|_\op\Big(
		\frac{r^2 \cdot
			\coher_U \cdot \coher_V}
			{d_1d_2}\Big)^{1/2}
		\, ,\\
		\|X^\sT e_i\|_2
		&=
			\|X^\sT UU^\sT e_i\|_2
			\le \|X\|_\op \|UU^\sT e_i\|_2
			\le \|X\|_\op 
			\Big(\frac{r\cdot \coher_U}
				{d_1}\Big)^{1/2}\, ,\\
		\|X e_\ell\|_2
		&=\|X VV^\sT e_\ell\|_2
		\le \|X\|_\op \|VV^\sT e_\ell\|_2
		\le \|X\|_\op 
		\Big(\frac{r \cdot \coher_V}
			{d_2}\Big)^{1/2}\,,
	\end{align*}
which proves the claim.
\end{proof}
\end{lem}

\begin{lem}\label{lemma:WLOG_Coh}
For any $X\in\R^{d_1\times d_2}$,
the parameters $\LL,\MM,\RR$
of Assumption~\ref{d:incoherent}
 can be chosen so that
	\beq\label{e:param.bounds}
	1/d_1\le\LL\le d_1, \
	1/d_2\le\RR\le d_2, \
	1\le \LL\MM\RR \le d_1d_2\,.\eeq
\begin{proof}
The quantities
	$\|X^\sT e_i\|$, $\|X e_\ell\|$,
 $|X_{i\ell}|$ 
are all trivially upper bounded by
$\|X\|_\op$, so we can always satisfy 
\ref{e:incoherent.rows},
\ref{e:incoherent.cols},
\ref{e:incoherent.max}
with $\LL\le d_1$, $\RR\le d_2$,
and $\LL\MM\RR \le d_1d_2$.
On the other hand
	\[\|X\|_\op
	=\bigg\|\sum_{\ell\le d_2}
	Xe_\ell (e_\ell)^\sT\bigg\|_\op
	\le \sum_{\ell\le d_2}
		\|X e_\ell\|_2
	\le \|X\|_\op (d_2\RR)^{1/2}\,,\]
which implies that \ref{e:incoherent.cols} can only be satisfied with $d_2\RR\ge1$,
and likewise 
\ref{e:incoherent.rows}
can only be satisfied with 
$d_1\LL\ge1$. Lastly, we have
	\[\|X\|_\op\le \|X\|_{\F}
	\le (d_1d_2)^{1/2}
		\max_{i,\ell}|X_{i,\ell}|
	\le 
	\|X\|_\op (\LL\MM\RR)^{1/2}\,,\]
so \ref{e:incoherent.max} can only be satisfied with $\LL\MM\RR\ge1$.
This concludes the justification of \eqref{e:param.bounds}.
\end{proof}
\end{lem}

\subsection{Proof of matrix estimation results}

We now prove Theorem~\ref{t:op.error}. Recall
that we assume a (deterministic) matrix
 $X\in\R^{d_1\times d_2}$, each entry of which is observed independently with chance $\delta\equiv \nn/(d_1d_2)$. Let $\ee\subseteq[d_1]\times[d_2]$ denote the subset of observed entries, and $Y=\Proj_\ee X$ the partially observed matrix. 
Let $I_{i\ell}$ be the indicator that $(i,\ell)$ belongs to the (random) set $\ee$; thus the $I_{i\ell}$ are i.i.d.\ $\mathrm{Ber}(\delta)$ random variables and $Y_{i\ell}
	=X_{i\ell} I_{i\ell}$.
As in \eqref{e:def.hB}, let
	\[\hB = \f1\delta\Proj(YY^\sT)
	+ \f1{\delta^2}\Proj_\perp(YY^\sT)\,.\]

\begin{proof}[Proof of Theorem~\ref{t:op.error}] 
We first make a preliminary remark that
	\beq\label{e:bound.delta}
	1\le\f{1}{\delta^2}
	\le \f{(d_1\LL) (d_2\RR)}{(\LL\MM\RR)^2}
	 \le (d_1d_2)^2\,.
	\eeq
(The second inequality follows from the assumptions, while the third follows from Lemma~\ref{lemma:WLOG_Coh}.)
We shall apply Proposition~\ref{p:trunc.bernstein} to bound
the spectral norm of
	\beq\label{e:def.Z.ell}
	\hB-XX^\sT
	=\hB-\E\hB
	=\sum_{\ell\le d_2} (B_\ell-\E B_\ell)
	=\sum_{\ell\le d_2} Z_\ell\,,
	\eeq
where $B_\ell$ is $d_1\times d_1$ with entries
	\[
	(B_\ell)_{ij}
	=\left\{\hspace{-4pt}\begin{array}{ll}
	\delta^{-1}
	(X_{i\ell})^2 I_{i\ell}
	&\textup{for }i=j\,,\\
	\delta^{-2}
	X_{i\ell}X_{j\ell}I_{i\ell}I_{j\ell},
	&\textup{for } i\ne j\,.\\
	\end{array}\right.
	\]
Lemmas~\ref{l:variance}~and~\ref{l:truncation} (below) show that the matrices $Z_\ell$ satisfy the hypotheses of Proposition~\ref{p:trunc.bernstein} with $\sigma^2$ as in \eqref{e:l.variance.bound} and $\beta,p,q$ as in \eqref{e:beta.p.q}, for $d_1d_2$ sufficiently large. We then have
	\[3\min\bigg\{\f{t^2}{\sigma^2},\f{t}{\beta}
	\bigg\}
	\ge \log(d_1d_2)^2\]
provided $t\ge t_{\max}
	\equiv (\log(d_1d_2))^4
	\max\set{t_1,t_2,t_3,t_4}\|X\|_\op^2$
for
	\begin{align*}
	t_1 &= \bigg(\f{\LL \RR}{d_1d_2\delta^2}\bigg)^{1/2}
	=\bigg(\f{\LL}{d_1 \delta}
		\bigg)^{1/2}
	\bigg(\f{\RR}{d_2\delta}\bigg)^{1/2}\,,\\
	t_2 
	&=\bigg(\f{\RR}{d_2\delta}\bigg)^{1/2}
	= t_1
	\bigg( \f{d_1\delta}{\LL} \bigg)^{1/2}\,,\\
	t_3
	&= \f{\LL\MM\RR}{d_1 d_2\delta^2}
	= \bigg(\f{\LL\RR}{d_1 d_2\delta^2}\bigg)^{1/2}
	\bigg(\f{\LL\MM^2\RR}{d_1 d_2\delta^2}\bigg)^{1/2}
	= t_1 \bigg(\f{\LL\MM^2\RR}{d_1 d_2\delta^2}\bigg)^{1/2}\,,\\
	t_4
	&= \f{(\LL\MM)^{1/2}\RR}
		{(d_1)^{1/2}d_2\delta}
	= \bigg(
	\f{\LL\RR}{d_1d_2\delta^2}
	\bigg)^{1/2}
	\bigg(\f{\MM\RR}{d_2}\bigg)^{1/2}
	= t_1
	\bigg(\f{\MM\RR}{d_2}\bigg)^{1/2}\,.
	\end{align*}
From \eqref{e:beta.p.q}
we have $t_{\max}\ge \beta_\star \ge d_2 q$. It follows from Proposition~\ref{p:trunc.bernstein}
 that 
	\[
	\P(\|\hB-\E\hB\|_\op \ge 2t_{\max})
	\le \f{2d_1d_2}{\exp\{
		\tfrac38
			\log(d_1d_2)^2 \}}
	+ \f{2d_1}{\exp\{ \tfrac18
		\log(d_1d_2)^2 \}}
	\le \f{4d_1 }{\exp\{ \tfrac18
		\log(d_1d_2)^2 \}}\,,
	\]
which concludes the proof.
\end{proof}

\begin{lem}\label{l:variance}
Assume the setting and notation of Theorem~\ref{t:op.error}, and let $Z_\ell$
be as defined by \eqref{e:def.Z.ell}.
For $\sigma^2$ as defined by 
\eqref{e:bernstein.rect.R.sigma}, we have
	\beq\label{e:l.variance.bound}
	\sigma^2 \le
	2\max\bigg\{
	\f{\LL\RR}{d_1d_2\delta^2},
	\f{\RR}{d_2\delta}
	\bigg\}
	\|X\|_\op^4\,.
	\eeq 

\begin{proof}
From the definitions, we have
	\[
	(Z_\ell)_{ij}
	=\left\{ \hspace{-4pt} \begin{array}{ll}
	\delta^{-1} (X_{i\ell})^2 (I_{i\ell}-\delta)
	& \textup{for } i=j\\
	\delta^{-2} X_{i\ell} X_{j\ell}
		(I_{i\ell} I_{j\ell}-\delta^2)
	& \textup{for } i\ne j\,.
	\end{array}\right.
	\]
Recalling \eqref{e:bernstein.rect.R.sigma}, let  $\Sigma$ denote the sum of the matrices
$\E[(Z_\ell)^2]$ over $\ell\le d_2$.
Let $W$ denote the $d_2\times d_2$
diagonal matrix with entries
$W_{\ell\ell} = \|X e_\ell\|^2$.
It is straightforward to compute that 
	\[
	\Sigma = 
	\f{1-\delta}{\delta^2} D
	+ \f{1-\delta}{\delta}
		X W X^\sT
	\]
where $D$ is the $d_1\times d_1$
diagonal matrix with entries
	\[
	D_{ii} = \sum_{\ell\le d_2}
		(X_{i\ell})^2
		\Big(\|Xe_\ell\|^2
		- (X_{i\ell})^2\Big)\,.\]
We then note $\|XWX^\sT\|_\op \le \|X\|_\op^2 \|W\|_\op \le \RR \|X\|_\op^4/d_2$, while
	\[
	0\le D_{ii}
	\le \f{\RR \|X\|_\op^2 \|X^\sT e_i\|^2}
		{d_2}
	\le \f{\LL\RR \|X\|_\op^4 }
		{d_1d_2}\,.
	\]
Combining the above estimates, we find
	\[
	\|\Sigma\|_\op
	\le
	\f{\|D\|_\op}{\delta^2}
	+ \f{\|XWX^\sT\|_\op}{\delta}
	\le 
	\f{\LL\RR\|X\|_\op^4 }{d_1d_2\delta^2}
	+ \f{\RR\|X\|_\op^4}{d_2\delta} \,,
	\]
yielding the claimed bound.
\end{proof}
\end{lem}

\begin{lem}\label{l:truncation} 
Assume the setting and notation of Theorem~\ref{t:op.error}, and let $Z_\ell$ be as defined by \eqref{e:def.Z.ell}. Let
	\[\beta_\star
	\equiv\max\bigg\{
	\f{
		\LL\MM\RR
		}{d_1 d_2\delta^2}
		\,,
	\f{(\LL\MM)^{1/2}\RR}
		{(d_1)^{1/2}d_2\delta}
	\bigg\}
	\|X\|_\op^2\,.
	\]
For $d_1d_2\ge3000$, the matrices $Z_\ell$ satisfy \eqref{e:bernstein.truncation} with
	\beq\label{e:beta.p.q}
	\begin{array}{rl}
	\beta 
	&\hspace{-6pt}
	=(\log(d_1d_2))^2 \beta_\star \,,\\
	p
	&\hspace{-6pt}=
	\exp\{-(3/8)(\log(d_1d_2))^2
	\}
	\cdot 2d_1\,,\\
	q 
	&\hspace{-6pt}=
	\exp\{-(1/8)
	(\log(d_1d_2))^2\}
	\cdot \beta_\star\,.
	\end{array}\eeq

\begin{proof}
Write $e_i$ for the $i$-th standard basis vector in $\R^{d_1}$, and let $E_{ij}\equiv e_i(e_j)^\sT$. Then $Z_\ell$ is the sum of independent zero-mean matrices $M_{ij}=E_{ii} (Z_\ell)_{ii}$. It is straightforward to calculate that It follows from the matrix Bernstein inequality (Proposition~\ref{p:bernstein.rect}) that
	\[
	\P(\|Z_\ell\|_\op\ge t)
	\le 2 d_1
	\exp\bigg\{ -\f38
		\min\bigg\{ 
		\f{t^2}{\sigma^2},\f{t}{R}
		\bigg\} \bigg\}
	\]
where (cf.~\eqref{e:bernstein.rect.R.sigma})
$R,\sigma$ are given by
	\begin{align*}
	\sigma
	&=\f{\|X\|_\infty\|X e_\ell\|}{\delta}
	\le 
	\f{(\LL\MM)^{1/2}\RR}
		{ (d_1)^{1/2} d_2}
	\f{\|X\|_\op^2}{\delta}
	\le
	d_1d_2 \|X\|_\op^2\,,\\
	R&=\f{\|X\|_\infty^2}{\delta^2}
		\le
		\f{
		\LL\MM\RR
		}{d_1 d_2}
		\f{\|X\|_\op^2}{\delta^2}
	\le
	(d_1d_2)^2 \|X\|_\op^2\,,
	\end{align*}
having made use of
Lemma~\ref{lemma:WLOG_Coh} and \eqref{e:bound.delta}.  
If we set 
$\beta_\star=\max\set{\sigma,R}$
and
$\beta =
(\log(d_1d_1))^2 \beta_\star$, then
	\[
	\P(\|Z_\ell\|_\op\ge \beta)
	\le 2 d_1
	\exp\{-\tfrac38
	(\log(d_1d_1))^2\}
	=p\,.
	\]
Next note that
for $t\ge\max\set{\sigma,R}$ we have
 $\min\set{t^2/\sigma^2,t/R}
 \ge t/\max\set{\sigma,R}$, so
	\begin{align*}
	&\E[\|Z_\ell\|_\op
	;\|Z_\ell\|_\op\ge \beta]
	\le \beta p
		+ \int_\beta^\infty
			\P(\|Z_\ell\|_\op\ge t)\,dt
	\le \beta p
		+ 2d_1
		\int_\beta^\infty
			\exp\bigg\{ -\f{(3/8) t}
				{\max\set{\sigma,R}}
				\bigg\}
			\,dt \\
	&\qquad=\f{
	[(\log(d_1d_2))^2 
	+ \tfrac83]2d_1
		\max\set{R,\sigma}}
		{ \exp\{\tfrac38(\log(d_1d_2))^2\} }
	\le
	\f{(d_1d_2)^2
	\max\set{R,\sigma} }
	{ \exp\{\tfrac38(\log(d_1d_2))^2\} }
	\le
	\f{\max\set{R,\sigma}}{ \exp\{
		\tfrac18 (\log(d_1d_2))^2\} }
	=q\,.
	\end{align*}
This concludes the proof.
\end{proof}
\end{lem}

\section{Tensor completion via unfolding}\label{appx:tensor.unfold}

In this section we prove Theorem~\ref{t:tensor}.
In the original model, we observe 
exactly $\delta=|\ee|/d^k$ fraction of the entries, uniformly at random. For convenience we now introduce the Bernoulli model where
each entry is observed independently with chance $\delta$. Our results for the Bernoulli model transfer to the original model by a standard argument, which we provide below.

As in Theorem~\ref{t:tensor},
suppose 
$\bT\in(\R^d)^{\otimes k}$ is a 
deterministic symmetric tensor
satisfying Assumption~\ref{ass:Main}
with unfolding parameters
$(\RU,\ALPHA,\MU)$.
Fixing $\delta\in(0,1)$, let $\delta_1\equiv\delta/2$
and $\delta_2\equiv\delta_1/(1-\delta_1)$.
Let $\set{I_\uu,J_\uu}$ be a collection of independent random variables (indexed by $\uu\in[d]^k$)
with $I_\uu\sim\Bern(\delta_1)$
and $J_\uu\sim\Bern(\delta_2)$.
Let $\ee_1$ be the set of $\uu\in[d]^k$ with $I_\uu=1$;
and let
$\ee_2$ be the set of
 $\uu\in[d]^k$ with $(1-I_\uu)J_\uu=1$.
Define the corresponding partially observed tensors $\bY_i=\Pi_{\ee_i}(\bT)$.
Fixing integers $1\le a \le b=k-a$, let
$X=\unfold^{a\times b}(\bT)$, $Z=\unfold^{a\times b}(\bY_1)$, and (cf.\ \eqref{eq:Bdef})
	\beq\label{eq:Bdef.subsample}
	B = \f{1}{\delta_1}\Proj(ZZ^\sT)
	+\f{1}{(\delta_1)^2}\Proj_\perp(ZZ^\sT).
	\eeq
Let $Q$ be the orthogonal projection onto the span of onto the eigenspace of $B$ for eigenvalues $\ge\lmstar$.
If $a=\lfloor k/2\rfloor$, then we can use $Q$ to define $\cQ$ as in \eqref{e:tensor.right.Q}. Then let
    \beq\label{e:repeat.def.hbTstar}
	\hbT\equiv
	\bY_1+\f1{\delta_2}\bY_2,\quad
	\bTstar=\cQ(\hT),\quad
	\hbTstar=\cQ(\hbT)\,;
	\eeq
and note that $\E[\hbT\,|\,\ee_1]=\bT$. Define
	\beq\label{e:DEF.TENSOR.QTYS}
	\varpi\equiv\f{\ALPHA\RU}{\MU}\,,\quad
	\THETA\equiv\f{\ALPHA\RU}{\delta d^{k/2}}\,,\quad
	\ETA(t)\equiv\f{8(k\log d)^4 t^{1/2}\varpi}
		{d^{k/2}\delta}\,,\quad
	\lmstar(t)\equiv\ETA^{2/3}\|B\|_\op\,.
	\eeq
We will consider $\hbTstar$ with threshold $\lmstar=\lmstar(t)$ as given by \eqref{e:DEF.TENSOR.QTYS}.

\begin{thm}\label{t:tensor.bernoulli} 
Suppose $\bT\in(\R^d)^{\otimes k}$ is a deterministic symmetric tensor satisfying Assumption~\ref{ass:Main}
with unfolding parameters
$(\RU,\ALPHA,\MU)$,
such that $\RU\le r_{\max}$
as defined by \eqref{e:r.max}. Fix $t\ge1$ and suppose
	\beq\label{e:TENSOR.DELTA.BDS}
	\f{32(k\log d)^4 t^{1/2}\varpi}
		{d^{k/2}} \le
	\delta \le \f{t\varpi}{d^a}
	\eeq
Then, with $\smash{\hbTstar}$ as above, we have
    \[\|\bT-\hbTstar\|_\F\le 20(k\log d)^{4/3}
	(t\MU)^{1/6}\THETA^{1/3}\|\bT\|_\F\]
with probability at least 
$1- 3 d^k \exp\{ -\tfrac18(k\log d)^2 \}$.
\end{thm}

Let us discuss the choice of $t$ in Theorem~\ref{t:tensor.bernoulli}. We wish to have a small error $\|\bT-\hbTstar\|_\F$, while ensuring that condition~\eqref{e:TENSOR.DELTA.BDS} is satisfied. First note that \eqref{e:TENSOR.DELTA.BDS} cannot be satisifed at all unless we have $ t^{1/2} \ge 32(k\log d)^4/d^{k/2-a}$.
If we take $\epsilon\le1$ and set
	\[
	\delta = \f{20^3 (k\log d)^4 t^{1/2}
		\MU^{3/2}\varpi}{\epsilon^3 d^{k/2}}\,,
	\]
then Theorem~\ref{t:tensor.bernoulli} 
gives $\|\bT-\hbTstar\|_\F\le\epsilon\|\bT\|_\F$.
This choice of $\delta$ automatically satisfies the lower bound of \eqref{e:TENSOR.DELTA.BDS}.
To satisfy the upper bound we require
	\[
	t^{1/2}
	\ge \f{20^3(k\log d)^4 \MU^{3/2}}
		{ \epsilon^3 d^{k/2-a} }\,.
	\]
Since $a\le k/2$ and we aim for 
$\epsilon \le 1/(k\log d)$ in the worse case, we
shall set
$t^{1/2} = (k\log d)^8 \MU^{3/2}$.
With this choice, \eqref{e:TENSOR.DELTA.BDS} simplifies to
	\[\f{32(k\log d)^{12}\ALPHA\RU \MU^{1/2}}{d^{k/2}}
	\le \delta
	\le \f{(k\log d)^{16} \ALPHA\RU\MU^2}{d^a}\,,
	\]
and we obtain $\|\bT-\hbTstar\|_\F\le\epsilon\|\bT\|_\F$ with
	\[\lmstar=
		4(k\log d)^8
		\bigg(
		\f{\ALPHA\RU\MU^{1/2}}{d^{k/2}\delta}
		\bigg)^{2/3}
		\|B\|_\op\,,\quad
	\epsilon
	= 20(k\log d)^3 \bigg(
		\f{\ALPHA\RU \MU^2}{d^{k/2}\delta}
		\bigg)^{1/3}\,.
	\]
Then, as noted previously,
the result of Theorem~\ref{t:tensor.bernoulli}
(for the Bernoulli model)
implies the result of
Theorem~\ref{t:tensor}
(for the original model)
by a well-known argument:

\begin{proof}[Proof of Theorem~\ref{t:tensor}]
The bound of Theorem~\ref{t:tensor.bernoulli} fails with probability tending to zero more rapidly than any polynomial of $d$. On the other hand, by construction,
the probability of the event
$|\ee_1|=|\ee_2|=\nn/2$ is lower bounded by a polynomial in $\nn$, so the result follows.
\end{proof}

\subsection{Preliminary lemmas}

We begin with a proof of our earlier remark~\eqref{e:r.ubd.tucker.rank}; note however that this bound is not used in the proof of Theorems~\ref{t:tensor}~or~\ref{t:tensor.bernoulli}.

\begin{lem}\label{l:tucker}
Suppose that the tensor $\bT\in\R^{d_1}\otimes\cdots\otimes\R^{d_k}$
has rank $r=r(\bT)$,
and multilinear rank
$\rmum=\rmum(\bT)$
--- recalling \eqref{e:tucker},
$\rmum$ is the maximum of the values
 $\rmu{i}=\rmu{i}(\bT)$
 over $i\le k$. Then (cf.\ \eqref{e:r.ubd.tucker.rank})
	\[
	r\le (\rmu{1} \cdots \rmu{k})/(\rmum)
	\le (\rmum)^{k-1}\,.
	\]
\begin{proof}
If $k=2$ this is clear from the singular value decomposition of the $d_1\times d_2$ matrix $\bT$. For $k\ge3$ we argue by induction on $k$.
By relabelling, we can suppose without loss
of generality that
$\rmum
\in\set{\rmu{2},\ldots,\rmu{k}}$.
Take a singular value decomposition
$X^{(1)} = UV^\sT$, where $U$ is a $d_1\times \rmu{1}$ matrix whose columns
form an orthonormal basis of
the space $\spn^{(1)}(\bT)$.
Column $s$ of $V$ defines a tensor
$\bV_s$, and likewise row $j$ of $X^{(1)}$ defines a tensor $\bT_j$ ---
both $\bV_s,\bT_j$
lie in
$\R^{d_2}\otimes \cdots \otimes \R^{d_k}$.
Since $V^\sT=U^\sT X^{(1)}$, each $\bV_s$ is a linear combination of the tensors $\bT_j$.
It is clear that $\spn^{(i-1)}(\bT_j)\subseteq\spn^{(i)}(\bT)$ for every $j$, so $\spn^{(i-1)}(\bV_s)
\subseteq\spn^{(i)}(\bT)$ for every $s$.
This proves $\rmu{i-1}(\bV_s)\le\rmu{i}$.
By the inductive hypothesis,
together with the assumption
$\rmum\in\set{\rmu{2},\ldots,\rmu{k}}$, we have
	\[
	r(\bV_s)
	\le (\rmu{2}\cdots\rmu{k})/(\rmum)\,.
	\]
It follows from the decomposition
$X^{(1)}=UV^\sT$ that
	\[
	r(\bT) \le
	\rmu{1}
	\max_s r(\bV_s)
	\le (\rmu{1}\cdots\rmu{k})/(\rmum)\,,
	\]
which verifies the inductive hypothesis and proves the claim.
\end{proof}
\end{lem}

The remainder of this section is devoted to the proof of Theorem~\ref{t:tensor.bernoulli}.

\begin{lem}\label{l:rank.Q}
If $A_1,A_2\in\R^{d\times d}$ with
$\|A_1-A_2\|_\op<1$,
and $A_2$ is an orthogonal projection matrix,
then $\rank A_2\le \rank A_1$.

\begin{proof}
Suppose $\rank A_2=r$, and take an orthogonal set of vectors
$x_1,\ldots,x_r\in\R^d$
with $A_2 x_j = x_j$ for all $j\le r$.
We claim that the vectors $A_1x_j$ are linearly independent --- to see this, suppose for contradiction that there exist constants $c_j$, not all zero,
such that the vector
	\[
	v = \sum_{j\le r} c_j x_j
	\]
lies in the kernel of $A_1$. Then
$v = (A_2-A_1)v$, so $\|v\|\le\|A_1-A_2\|_\op\|v\|$.
Since $\|A_1-A_2\|_\op<1$,
it follows that $v=0$, a contradiction.
It follows that the vectors $A_1x_j$
 are linearly independent, which proves 
 $\rank A_2=r\le \rank A_1$ as claimed. 
\end{proof}
\end{lem}

\begin{lem}\label{l:tensor.apply.concentration} Let $\bT\in(\R^d)^{\otimes k}$ be a deterministic tensor, not necessarily symmetric. Fixing integers $1\le a\le b=k-a$, let $X = \unfold^{a\times b}(\bT)$, and take $B$ as in \eqref{eq:Bdef.subsample}. Suppose $d^k\|X\|_\infty^2 \le \bar{\varpi}\|X\|_\op^2$ for some $\bar{\varpi}\ge1$.
For $t\ge1$, in the regime $1/(td^k)^{1/2} \le \delta \le t\bar{\varpi}/d^a$
we have
	\[\f{\|B-XX^\sT\|_\op}{8(k\log d)^4}
    \le\f{t^{1/2} \bar{\varpi} \|X\|_\op^2}{d^{k/2}\delta}
    \]
with probability at least 
$1-d^k\exp\{ -\tfrac18(k\log d)^2 \}$.

\begin{proof} Recalling \eqref{e:coherence.params.reln.if.bdd}, the matrix $X$ is $(\LL,\MM,\RR)$-incoherent with $\LL=\RR =1/\MM=\bar{\varpi}$. The claim then follows by applying Corollary~\ref{c:op.error.simp} (an additional factor $4$ in the bound arises since $\delta=2\delta_1$).
\end{proof}
\end{lem}

\begin{lem}\label{l:tensor.conc}
Suppose $F$ is a $d_1\times d_2$ matrix whose entries $F_{i,\ell}$ are independent random variables which have mean zero, variance at most 
$\nu^2$, and magnitude at most $R$ almost surely. Suppose we also have deterministic square matrices $A_1$ and $A_2$,
of dimensions $d_1$ and $d_2$ respectively,
with $\|A_i\|_\op\le1$. Then, for $t\ge0$, 
we have
	\[
	\|A_1F(A_2)^\sT\|_\op
	\le \max\Big\{
	\Big(t
	\nu^2 \max\set{\rank A_1,\rank A_2}
	\Big)^{1/2} \,, tR
	\Big\}
	\]
with probability at least
$1-(d_1+d_2)\exp\{-\tfrac38t\}$.

\begin{proof}
We can decompose 
	\[
	A_1F(A_2)^\sT
	=A_1\bigg( 
	\sum_{i,\ell}
	F_{i,\ell}
		e_i (e_\ell)^t \bigg) (A_2)^\sT
	=\sum_{i,\ell} Z_{i,\ell}
	\]
where $Z_{i,\ell}=F_{i,\ell} (A_1e_i)(A_2e_\ell)^\sT$ is a  $d_1\times d_2$ matrix.
It holds almost surely that
$\|Z_{i,\ell}\|_\op
\le |F_{i,\ell}|\le R$.
We also have the variance bounds
	\begin{align*}
	\bigg\|\sum_{i,\ell}
	\E[ Z_{i,\ell}(Z_{i,\ell})^\sT]
	\bigg\|_\op
	&\le 
	\bigg\|
	\sum_i
	A_1e_i(A_1e_i)^\sT
	\sum_\ell
	\E[ (F_{i,\ell})^2]
	((A_2)^\sT A_2)_{\ell,\ell}
	\bigg\|_\op\\
	&\le\nu^2 \trace ((A_2)^\sT A_2)
	\|A_1(A_1)^\sT\|_\op
	= \nu^2 \|A_2\|_\F^2 \|A_1\|_\op^2
	\le \nu^2 (\rank A_2)\,,
	\end{align*}
and in a symmetric manner
	\begin{align*}
	\bigg\|\sum_{i,\ell}
	\E[(Z_{i,\ell})^\sT Z_{i,\ell}]
	\bigg\|_\op
	\le \nu^2 (\rank A_1)\,.
	\end{align*}
The claimed bound follows by the matrix Bernstein inequality (Proposition~\ref{p:bernstein.rect}).
\end{proof}
\end{lem}

\subsection{Projection of original tensor}

Recalling
\eqref{e:tensor.right.Q}~and~\eqref{e:repeat.def.hbTstar},
we now compare
 the original tensor $\bT$ and its projection $\bTstar=\cQ(\bT)$.

\begin{lem}\label{l:tensor.rk.bound}
Let $\bT\in(\R^d)^{\otimes k}$
be a tensor (not necessarily symmetric).
Fix integers $a,b\ge1$ with $a+b=k$, and let
$X=\unfold^{a\times b}(\bT)$. For any positive semidefinite matrix $B$
of dimension $d^a$,
let $Q$
be the orthogonal projection onto the eigenspace of $B$  corresponding to eigenvalues $\ge\lmstar$. Then, for any $\lmstar>\|B-XX^\sT\|_\op$,
	\[\rank Q\le\rank X\,.\]
In particular, if $a=\lfloor k/2\rfloor$
then $\rank\cQ\le r_\circ(k,\rank X,d)$ where
(cf.\ \eqref{e:tensor.right.Q})
	\beq\label{e:tensor.right.Q.repeat}
	\cQ=\left\{\hspace{-4pt}\begin{array}{ll}
	Q\otimes Q\otimes Q, & k=3,\\
	Q\otimes Q, & k\ge4\textup{ even},\\
	Q\otimes Q \otimes \id_d
		, & k\ge5\textup{ odd},
	\end{array}\right.
	\qquad
	r_\circ(k,r,d)
	\equiv\left\{\hspace{-4pt}\begin{array}{ll}
	r^3, & k=3,\\
	r^2, & k\ge4\textup{ even},\\
	r^2 d, & k\ge5\textup{ odd}.
	\end{array}\right.
	\eeq

\begin{proof}
Let $P$ be the orthogonal projection onto the eigenspace of $XX^\sT$ corresponding to eigenvalues $\ge2\lmstar$;
and note $\rank(PQ)
\le \rank P \le \rank X$.
From Wedin's theorem (Proposition~\ref{p:wedin}),
	\beq\label{e:tensor.apply.wedin}
	\|P(I-Q)\|_\op
	=\|(I-Q)P\|_\op
	\le\f{\|B-XX^\sT\|_\op}{\lmstar}\,,
	\eeq
which is less than one by assumption.
Applying Lemma~\ref{l:rank.Q} then gives 
\[\rank Q\le \rank(PQ)\le \rank X\,,\]
proving the first assertion.
The claimed bound on $\rank\cQ$
 follows immediately
 from the fact that $\rank(M_1\otimes M_2) = \rank(M_1)\rank(M_2)$.
\end{proof}
\end{lem}

\begin{lem}\label{l:T.replace}
Let $\bT\in(\R^d)^{\otimes k}$ be a  symmetric tensor. Take $a=\lfloor k/2\rfloor$ and let $X,B,Q,\cQ$
 be as in the statement of Lemma~\ref{l:tensor.rk.bound}.
Then $\bTstar=\cQ(\bT)$ satisfies
	\[
	\|\bT-\bTstar\|_\F
	\le 3 (\rank X)^{1/2}\bigg(
		(2\lmstar)^{1/2}
		+\f{
		\|B-XX^\sT\|_\op
		\|X\|_\op}{\lmstar}
		\bigg)\,.
	\]

\begin{proof}
In what follows we write $\id$ for the $d\times d$ identity matrix. We denote its $\ell$-fold tensor product by $\id^{(\ell)}\equiv\id^{\otimes\ell}$; this is equivalent
to the $d^\ell\times d^\ell$ identity matrix.
With this notation we expand
	\[
	\bT-\bTstar
	= \bigg((\id^{(a)}-Q)
		\otimes \id^{(b)}
	+ Q\otimes(\id^{(a)}-Q)
		\otimes \id^{(b-a)} 
	+\Ind{k=3} 
		Q\otimes Q\otimes (\id^{(a)}-Q)
		\bigg)
		\bT\,.
	\]
Recall $X=\unfold^{a\times b}(\bT)$.
By the triangle inequality and the assumed symmetry of $\bT$, we have
	\[
	\|\bT-\bTstar\|_\F
	\le 3\max_M\Big\{\|
		(\id^{(a)}-Q)XM
		\|_\F\Big\}
	\]
where the maximum is taken over all $d^b\times d^b$ matrices $M$ with $\|M\|_\op\le1$. 
Then, with $P$ as in the proof of
Lemma~\ref{l:tensor.rk.bound}, we can expand,
	\[
	(\id^{(a)}-Q)XM
	= (\id^{(a)}-Q)
		(\id^{(a)}-P)XM
	+(\id^{(a)}-Q) P XM\,,
	\]
and bound separately the two terms on the right-hand side. For the first term we have
	\[
	\|(\id^{(a)}-Q)(\id^{(a)}-P)XM\|_\op
	\le \|(\id^{(a)}-P)X\|_\op
	\le (2\lmstar)^{1/2}\,,
	\]
from the definition of $P$.
For the second term we have
(cf.~\eqref{e:tensor.apply.wedin})
	\[
	\|(\id^{(a)}-Q) P XM\|_\op
	\le \|(\id^{(a)}-Q) P \|_\op\|X\|_\op
	\le \f{
		\|B-XX^\sT\|_\op
		\|X\|_\op}{\lmstar}\,.
	\]
Combining the above inequalities gives
	\[
	\|(\id^{(a)}-Q)XM\|_\op
	\le
	(2\lmstar)^{1/2}
	+ \f{
	\|B-XX^\sT\|_\op
	\|X\|_\op}{\lmstar}\,.
	\]
The claimed bound follows by noting that
the matrix $(\id^{(a)}-Q)XM$ has rank 
upper bounded by $\rank X$,
so its Frobenius norm is at most
$(\rank X)^{1/2}$
times its spectral norm.
\end{proof}
\end{lem}

In view of Lemma~\ref{l:T.replace}, it is natural to optimize over the parameter $\lmstar$
by setting
	\[
	\lmstar = 
	\bigg(
	\f{\|B-XX^\sT\|_\op^2\|X\|_\op^2}
	{ 2 }
	\bigg)^{1/3}\,.
	\]
Of course, in the application we have in mind, we cannot do this because $X$ is unknown. 
However, if the (known) matrix $B$ is sufficiently close to $XX^\sT$, we can achieve a near-optimal bound by defining $\lmstar$ in terms of $B$ alone, without reference to $X$. In summary, we have:

\begin{cor}\label{c:T.replace}
Suppose $\bT\in(\R^d)^{\otimes k}$ is a deterministic symmetric tensor satisfying
Assumption~\ref{ass:Main}.
Take $a=\lfloor k/2\rfloor$ and
define $B$ as in \eqref{eq:Bdef.subsample}. 
Recalling \eqref{e:DEF.TENSOR.QTYS}, let $Q$
be the orthogonal projection onto the eigenspace of $B$  corresponding to eigenvalues $\ge\lmstar(t)$, and use this to define $\bTstar=\cQ(\bT)$
as in \eqref{e:repeat.def.hbTstar}. For $t\ge1$
and $\delta$ satisfying \eqref{e:TENSOR.DELTA.BDS}, we have
	\[\|\bT-\bTstar\|_\F
	\le 18(k\log d)^{4/3}
	(t\MU)^{1/6}
	\vartheta^{1/3}
	\|\bT\|_\F\]
with probability at least 
$1-d^k\exp\{ -\tfrac18(k\log d)^2 \}$.

\begin{proof}
Since $t\ge1$ and $\varpi=\ALPHA\RU/\MU\ge1$
(Remark~\ref{r:WLOG.coherence.tensor}),
it follows from \eqref{e:TENSOR.DELTA.BDS} that
	\[
	\f{1}{(td^k)^{1/2}}
	\le \f{32(k\log d)^4 t^{1/2}\varpi}
		{d^{k/2}} \le
	\delta \le \f{t\varpi}{d^a}\,.
	\]
Together with \ref{e:assump.max.frob}~and~\ref{e:assump.frob.op},
we see that the conditions of Lemma~\ref{l:tensor.apply.concentration}
are satisfied with $\bar{\varpi}=\varpi$. 
It follows that,
with probability 
at least $1-d^k\exp\{-\tfrac18(k\log d)^2 \}$,
	\[\|B-XX^\sT\|_\op
	\le\ETA\|X\|_\op^2
	\le \tfrac14\|X\|_\op^2\,,
	\]
where the last inequality is from 
\eqref{e:TENSOR.DELTA.BDS}.
Therefore $\tfrac34\|X\|_\op^2\le \|B\|_\op \le \tfrac54\|X\|_\op^2$, so
	\beq\label{e:lambda.star.bounds}
	\tfrac34\ETA^{2/3} \|X\|_\op^2
	\le \lmstar(t)
	\le\tfrac54\ETA^{2/3} \|X\|_\op^2\,.\eeq
(So far, it was not necessary for $\bT$ to be symmetric.) Next, substituting \eqref{e:lambda.star.bounds}
into the bound of Lemma~\ref{l:T.replace}
(and making use of the symmetry of $\bT$)
we find
	\[
	\|\bT-\bTstar\|_\F
	\le 9\ETA(t)^{1/3}
	(\rank X)^{1/2}\|X\|_\op
	\le 9\ETA(t)^{1/3}
	\RU^{1/2}\|X\|_\op\,,
	\]
where the last inequality is from \ref{e:assump.rank}.
Finally, applying \ref{e:assump.frob.op}
and recalling $\|X\|_\F=\|\bT\|_\F$, we conclude
	$
	\|\bT-\bTstar\|_\F
	\le 9
	\MU^{1/2} \ETA(t)^{1/3} \|\bT\|_\F$.
The claim follows by recalling the 
definition of $\ETA(t)$ from \eqref{e:DEF.TENSOR.QTYS}.
\end{proof}
\end{cor}

\subsection{Projection of observed tensor}

Again recalling
\eqref{e:tensor.right.Q}~and~\eqref{e:repeat.def.hbTstar},
we next compare 
$\bTstar=\cQ(\bT)$,
(the projection of the original tensor)
with $\hbTstar=\cQ(\hbT)$
(the projection of the observed tensor).

\newcommand{\rankq}{\mathcal{R}}

\begin{lem}\label{l:split.sample.conc}
Let $\bT\in(\R^d)^{\otimes k}$ be a 
deterministic tensor (not necessarily symmetric).
Fix integers $a,b\ge1$ with $a+b=k$. Suppose
we have two $\ee_1$-measurable square matrices
$A_1$ and $A_2$,
of dimensions $d^a$ and $d^b$ respectively,
with $\|A_i\|_\op\le1$.
Let $\cQ\equiv A_1\otimes A_2$,
and abbreviate $\rankq=\max\set{\rank A_1,\rank A_2}$. For this choice of $\cQ$,
define $\smash{\bTstar}$ and $\smash{\hbTstar}$
as in \eqref{e:repeat.def.hbTstar}. Then
	\[
	\Big\|\unfold^{a\times b}
	(\bTstar-\hbTstar)
	\Big\|_\op
	\le2(k\log d)^2
	\bigg(
	\f{\max\set{\delta^{-1},\rankq}}
		{\delta} \bigg)^{1/2}
	\|\bT\|_\infty
	\]
with probability at least $1-d^k\exp\{-\tfrac38(k\log d)^2\}$
conditional on $\ee_1$.

\begin{proof} Let $\bF=\bT-\hbT$; it follows from the definitions that $\bF$ has entries
	\[F_\uu
	= (1-I_\uu) \Big(1-\f{J_\uu}{\delta_2}\Big) 
			T_\uu\,.\]
Note that $\E[F_\uu\,|\,\ee_1]=0$, or equivalently $\E[\hbT\,|\,\ee_1]=\bT$. Moreover we have
    \[\esssup|F_\uu| \le \f{\|\bT\|_\infty}{\delta_2},
    \quad
    \E[(F_\uu)^2 | \ee_1]
    \le \f{\|\bT\|_\infty^2}{\delta_2}.\]
If $F=\unfold^{a\times b}(\bF)$ then
we have
	\[\unfold^{a\times b}(\bTstar-\hbTstar)
	=\unfold^{a\times b}(\cQ(\bT-\hbT))
	= A_1 F (A_2)^t\,.\]
The claimed bound then follows from Lemma~\ref{l:tensor.conc}
(and using $2\delta_2\ge\delta$).
\end{proof}
\end{lem}

\begin{cor}\label{c:split.sample.conc} In the setting of Lemma~\ref{l:split.sample.conc}, suppose $\bT$ satisfies \ref{e:assump.max.frob}~and~\ref{e:assump.frob.op}, as well as
	\beq\label{e:lbd.delta.two}
	\max\set{\delta^{-1},\rankq}\le d^{k/2}\,.
	\eeq
Then, conditional on $\ee_1$, 
and with $\THETA\equiv
\ALPHA\RU/(d^{k/2}\delta)$, we have
	\[
	\Big\|\unfold^{a\times b}
	(\bTstar-\hbTstar)\Big\|_\op
	\le2
	(k\log d)^2
	(\THETA/\MU)^{1/2}
	\|X\|_\op
	\]
with probability at least $1-d^k\exp\{-\tfrac18(k\log d)^2\}$.

\begin{proof} 
From \ref{e:assump.max.frob}~and~\ref{e:assump.frob.op}
we have
	\[
	\f{d^{k/2}}{\delta}\|\bT\|_\infty^2
	\le
	\f{\ALPHA\RU}{\MU d^{k/2}\delta} \|\bT\|_\op^2
	=\f{\THETA}{\MU}\|\bT\|_\op^2\,.
	\]
Combining with~\eqref{e:lbd.delta.two}
and substituting
into Lemma~\ref{l:split.sample.conc}
gives the claim.
\end{proof}
\end{cor}

\begin{cor}\label{c:split.sample.conc.conclusion}
Let $\bT\in(\R^d)^{\otimes k}$ be a deterministic tensor (not necessary symmetric)
satisfying Assumption~\ref{ass:Main}
with unfolding parameters $(\RU,\ALPHA,\MU)$
with $\RU\le r_{\max}$ (as defined by \eqref{e:r.max}).
Fixing $t\ge1$, suppose $\delta$ satisfies \eqref{e:TENSOR.DELTA.BDS},
and set $\lmstar$ as in \eqref{e:DEF.TENSOR.QTYS}. 
With $\bTstar$ and $\smash{\hbTstar}$ as in 
\eqref{e:repeat.def.hbTstar}, we have 
	\[\|\bTstar-\hbTstar\|_\F
	\le 2(k\log d)^2\THETA^{1/2}
	\|\bT\|_\F\]
with probability at least
$1-2d^k\exp\{-\tfrac18(k\log d)^2\}$.
	
\begin{proof}
Fix $a=\lfloor k/2\rfloor$
and $b=\lceil k/2\rceil$. Recalling the proof of Corollary~\ref{c:T.replace},
with probability
at least $1-d^k\exp\{ -\tfrac18(k\log d)^2 \}$
the bounds
\eqref{e:lambda.star.bounds}
hold, in which case
Lemma~\ref{l:tensor.rk.bound} gives $\rank Q\le \rank X$.
We also have $\rank X\le\RU$ by~\ref{e:assump.rank}.
From  \eqref{e:tensor.right.Q}, $\cQ=A_1\otimes A_2$ where
$A_1=Q$ and 
	\[
	A_2=\left\{\hspace{-4pt}
	\begin{array}{ll}
	Q\otimes Q, & k=3,\\
	Q\otimes \id^{(b-a)}, & k\ge4.
	\end{array}\right.
	\]
As in the proof of Lemma~\ref{l:split.sample.conc} denote
$\bF=\bT-\hbT$,
and $F=\unfold^{a\times b}(\bF)$.
Then $\bTstar-\hbTstar=\cQ(\bF)$, and the matrix $\unfold^{a\times b}(\bTstar-\hbTstar) =A_1 F (A_2)^\sT$ has rank upper bounded by
the rank of $A_1=Q$. 
We have seen that with high probability $\rank Q\le\RU$ --- on this event,
	\[
	\|\bTstar-\hbTstar\|_\F
	\le \RU^{1/2}\|
	\unfold^{a\times b}(\bTstar-\hbTstar)\|_\op
	\,.\]
Condition~\eqref{e:lbd.delta.two}
is satisfied by our assumptions, so we can apply 
Corollary~\ref{c:split.sample.conc}:
conditional on $\ee_1$ it holds with probability
 $\ge1-d^k\exp\{  -\tfrac18(k\log d)^2\}$
that the right-hand side above is
	\[
	\le2(k\log d)^2\THETA^{1/2}
	\bigg(\f{ \RU^{1/2}\|X\|_\op}{\MU^{1/2}}
	\bigg)
	=2(k\log d)^2
	\THETA^{1/2}\|X\|_\F,
	\,.
	\]
where the last step uses \ref{e:assump.frob.op}. The claim follows since $\|X\|_\F=\|\bT\|_\F$.
\end{proof}
\end{cor} 

\begin{proof}[Proof of Theorem~\ref{t:tensor.bernoulli}]
The result now follows straightforwardly by collecting the estimates obtained above.
By our assumptions on $\delta$ and $\RU$,
the conditions of Corollaries~\ref{c:T.replace}~and~\ref{c:split.sample.conc.conclusion}
are satisfied. 
By  Corollary~\ref{c:T.replace},
it holds with probability at least $1-d^k\exp\{-\tfrac18(k\log d)^2\}$ that
	\[
	\|\bTstar-\hbTstar\|_\F
	\le 18(k\log d)^{4/3}
	(t\MU)^{1/6}
	\vartheta^{1/3}
	\|\bT\|_\F\,.
	\]
By Corollary~\ref{c:split.sample.conc.conclusion},
it holds with probability at least
$
1-2d^k\exp\{-\tfrac18(k\log d)^2\}$ that
	\[
	\|\bTstar-\hbTstar\|_\F
	\le 2(k\log d)^2\THETA^{1/2}
	\|\bT\|_\F\,.
	\]
Combining \eqref{e:DEF.TENSOR.QTYS}~with~\eqref{e:TENSOR.DELTA.BDS} gives
	\[
	\f{\THETA}{t\MU}
	=\f{\ALPHA\RU}{d^{k/2}\delta t\MU}
	=\f{\varpi}{d^{k/2}\delta t}
	\le \f1{32(k\log d)^4 t^{3/2}}\,.
	\]
Combining the above bounds gives
	\[
	\|\bT-\hbTstar\|_\F
	\le\|\bT-\bTstar\|_\F
	+\|\bTstar-\hbTstar\|_\F
	\le 20(k\log d)^{4/3}
	(t\MU)^{1/6}\THETA^{1/3}\|\bT\|_\F
	\]
which concludes the proof.
\end{proof}

\section{Overcomplete random three-tensors}
\label{appx:three}

In this section we prove Theorem~\ref{t:three}.
We have an underlying tensor
\beq\label{e:recall.three.tensor}
	\bT = \sum_{s\le r} 
	a_s\otimes a_s\otimes a_s
	\eeq
where $a_1,\ldots,a_r$ are i.i.d.\ random vectors in $\R^d$ satisfying
\ref{e:vector.symmetric},
\ref{e:vector.isometric}~and~
\ref{e:vector.subgaussian}.
We contract two copies of the tensor $\bT$ together to form the $d^2\times d^2$ matrix
$G$, with  entries
	\[
	G_{\ui,\uj}
	=\sum_{\ell\le d}
	\bT_{\ell i_1 j_1}
	\bT_{\ell i_2 j_2}.\]
Equivalently, writing $A_s\equiv a_s (a_s)^\sT$,
we have
	\beq\label{e:contract.tensor}
	G= \sum_{s,t\le r}
	\<a_s,a_t\>
	(a_s\otimes a_t)
	(a_s\otimes a_t)^\sT
	=\sum_{s,t\le r}
	\<a_s,a_t\>
	A_s \otimes A_t
	= G^\dg + G^\cross,
	\eeq
where $G^\dg$ denotes the contribution from the diagonal terms $s=t$, while $G^\cross$ denotes the remaining contribution from pairs $s\ne t$.

As in the proof of Theorem~\ref{t:tensor},
we work under a 
Bernoulli model for the partially observed tensor:
define three $d\times d\times d$ arrays
of i.i.d.\ $\Bern(\delta)$ random variables,
denoted $I,J,K$. 
Define 
$\dot{\bY} = \bT\odot I$,
$\ddot{\bY} = \bT\odot J$,
$\dddot{\bY} = \bT\odot K$.
The observed version of $G$ is
(cf.\ \eqref{e:def.contract.tensor})
	\beq\label{e:obs.contract.tensor}
	W_{\ui,\uj} 
	= \f{1}{\delta^2}
	\sum_{\ell\le d}
	\dot{\bY}_{\ell i_1 j_1}
	\ddot{\bY}_{\ell i_2 j_2}\,.
	\eeq
Take the singular vectors of $W$ with singular values at least $\lmstar$,  let $Q:\R^{d^2}\to\R^{d^2}$ be the orthogonal projection onto their span,
and let $\cQ=Q\otimes \id_d$.
Let $\hbT=\delta^{-1}\dddot{\bY}$,
and $\hbTstar=\cQ(\hbT)$.

Throughout this Appendix, we use $f(d,r)\lesssim g(d,r)$ if $f(d,r)\le (\log d)^Cg(d,r)$ for some constant $C$, and 
$f(d,r) \asymp g(d,r)$ if $f(d,r) \lesssim g(d,r)$ and $g(d,r)\lesssim f(d,r)$.

\begin{thm}\label{t:three.bernoulli} Let $\bT\in(\R^d)^{\otimes 3}$  be a standard random tensor \eqref{e:random.tensor} satisfying Assumption~\ref{ass:Random}. Suppose $\delta^2 \max\set{r,d}\ge1$, and take $\hbTstar$ as above with threshold parameter
(cf.\ \eqref{e:three.lambda.star.in.n})
	\[\lmstar=
	\bigg(
	\f{\max\set{1,r/d}}{d^{1/2}\delta}
	\bigg)^{4/5}\,.\]
Then it holds with very high probability that
(cf.\ \eqref{e:three.error.bound})
	\[  
	\|\bT-\hbTstar\|_\F
	\lesssim
	\bigg(
	\f{\max\set{1,r/d}}{d^{1/2}\delta}
	\bigg)^{1/5}
	\|\bT\|_\F\,.
	\]
\end{thm}

\begin{proof}[Proof of Theorem~\ref{t:three}]
Theorem~\ref{t:three}
is deduced from Theorem~\ref{t:three.bernoulli}
in the same way that
Theorem~\ref{t:tensor}
is deduced from 
Theorem~\ref{t:tensor.bernoulli}.
\end{proof}

\subsection{Preliminaries on random vectors}
\label{ss:subgaussian}

We now collect some basic estimates on random vectors satisfying condition~\ref{e:vector.subgaussian}, which we repeat here for convenience:
	\[\E\exp(\<x,v\>)
	\le \exp\bigg\{ \f{\tau^2\|v\|^2}{2d}
	\bigg\}
	\quad\text{for all }v\in\R^d\,.\]
Such vectors will be termed ``$(\tau^2/d)$-subgaussian.''

\begin{lem}\label{lem:subgaussiannorm}
Suppose $x$ is a random vector in $\R^d$ satisfying \ref{e:vector.subgaussian}.
Then
	\[
	\E\exp\bigg\{
	\f{\psi\|x\|^2}{2\tau^2} \bigg\}
	\le
	\f1{ (1 -\psi/d)^{d/2} }
	\le \exp\bigg\{ \f{\psi(1+\psi/d)}{2}
	\bigg\}\,,
	\]
where the first inequality holds for all $0\le\psi<d$,
and the second holds for all $0\le\psi\le d/2$.

\begin{proof}
Let $\xi$ be a standard gaussian random vector in $\R^d$ (with covariance $\E[\xi\xi^\sT]$ given by the $d\times d$ identity matrix $\id_d$). Applying \ref{e:vector.subgaussian} then gives, for $0\le\lambda< d/(2\tau^2)$,
	\[
	\E\exp\{ \lambda\|x\|^2 \}
	= \E\exp\{ (2\lambda)^{1/2}\<\xi,x\>\}
	\le 
	\E\exp\Big\{ \f{\lambda\tau^2\|\xi\|^2}{d}
		\Big\}
	=
	\bigg( 1-\f{2\lambda\tau^2}{d}
	\bigg)^{-d/2}\,,
	\]
which proves the first inequality by setting $\psi=2\lambda\tau^2$. Next note that for $0\le t \le 1/2$
we have $-\log(1-t) \le t(1+t)$. The second inequality then follows, with $t=\psi/d$.
\end{proof}
\end{lem}

\begin{lem}\label{l:lower.tail.bound}
Suppose $x$ is a random vector in $\R^d$ satisfying 
\ref{e:vector.isometric}~and~\ref{e:vector.subgaussian}. Then
	\[
	\P(\|x\|^2\ge\theta )
	\ge \f{1-\theta}{\tau^2}
		- \f{O(\log d)}{d^{1/2}}\,.
	\]
for any $0\le\theta\le1-1/(\log d)$,
and it holds
for any fixed $v\in\R^d$ that
	\[
	\P(d\<x,v\>^2
	\ge \|v\|^2/100) 
	\ge \f{1}{ 2 \tau^2 ( 8+3\log(\tau^2)) }\,.
	\]

\begin{proof}
Recall that if $Z$ is a non-negative random variable with finite mean, then
	\[
	\E[Z ; Z\ge L]
	= L\P(Z\ge L)
	+\int_L^\infty\P(Z\ge z)\,\de z\,.
	\]
Therefore, for any $0\le\theta\le L$ we can bound
	\begin{align}\nonumber
	\E Z
	&\le
	\theta\P(Z\le\theta)
	+L\P(\theta<Z<L)
	+\E[Z;Z\ge L]\\ \nonumber
	&= L-(L-\theta)\P(Z\le\theta)
	+\int_L^\infty\P(Z\ge z)\,\de z\\
	&\le
	L-(L-\theta)\P(Z\le\theta)
	+ \f{\E\exp (\lambda Z)}
	{\lambda \exp(\lambda L)}\,.
	\label{e:prelim.ubd.L.theta}
	\end{align}
Taking $\theta<\min\set{\E Z,L}$ and rearranging gives
	\beq\label{e:second.ubd.L.theta}
	\P(Z\le\theta)
	\le\f{L-\E Z}{L-\theta}
	+\f{\E\exp (\lambda Z)}
		{(L-\theta)\lambda \exp(\lambda L)}
	\le 1-\f{\E Z-\theta}{L}
	+\f{\E\exp (\lambda Z)}
		{(L-\theta)\lambda \exp(\lambda L)}
		\,.
	\eeq
Turning to the proof of the claim, we now take
 $Z =\|x\|^2$, so $\E Z=1$
 by \ref{e:vector.isometric}.
First, taking $\theta=L$ in 
\eqref{e:prelim.ubd.L.theta}
and applying Lemma~\ref{lem:subgaussiannorm} gives (for any $0\le \psi\le d/2$)
	\[
	1 \le L + 
	\f{2\tau^2}{\psi}
	\E \exp\bigg\{
	\f{\psi(\|x\|^2-L)}{2\tau^2} \bigg\}
	\le
	L+
	\f{2\tau^2}{\psi}
	\exp\bigg\{ \f{\psi( \tau^2(1 + \psi/d)  - L)}{2\tau^2} \bigg\}\,.
	\]
Setting $L = \tau^2(1+\psi/d)$ and rearranging gives
	\beq\label{e:lbd.tau}
	\tau^2
	\ge \f1{1 + \psi/d + 2/\psi}
	\ge 1 - O(d^{-1/2})\,,\eeq
where the last inequality is by optimizing over $0\le \psi\le d/2$. Next consider \eqref{e:second.ubd.L.theta},
where we again set $L = \tau^2(1+\psi/d)$
with $0\le \psi\le d/2$,
but now take $\theta \le 1-1/(\log d)$.
It follows from \eqref{e:lbd.tau} that
$(L-\theta)^{-1} \le O((\log d)/L)\le O((\log d)/\tau^2)$. Substituting into \eqref{e:second.ubd.L.theta} gives
	\[
	\P(\|x\|^2\le\theta)
	\le 1-\f{1-\theta}{L}
		+ \f{2\tau^2/\psi}{L-\theta}
	\le 1-\f{1-\theta}{\tau^2(1+\psi/d)}
		+ \f{O(\log d)}{\psi}
	\le  1 - \f{1-\theta}{\tau^2}
		+ \f{O(\log d)}{d^{1/2}}\,,
	\]
where the last step is by optimizing over $0\le \psi\le d/2$ as before. This proves the first claim.

For the second claim, note that \ref{e:vector.subgaussian}
implies that $\<x,v\>\in\R^1$ is
$\bar{\tau}^2$-subgaussian with
$\bar{\tau}^2 = \tau^2 \|v\|^2/d$.
We assume without loss of generality that $\|v\|^2=d$, so that $\<x,v\>$ is 
$(\tau^2/d)$-subgaussian.
We also have
$\E[\<x,v\>^2]=1$
by \ref{e:vector.isometric}.
Applying \eqref{e:second.ubd.L.theta} then gives
	\[
	\P(\<x,v\>^2\le\theta)
	\le 1-\f{1-\theta}{L}
		+
		\f{2\tau^2/\psi}{L-\theta}
		\E \exp\bigg\{
	\f{\psi(\<x,v\>^2-L)}{2\tau^2} \bigg\}\,.
	\]
Applying Lemma~\ref{lem:subgaussiannorm}
with $d=1$ gives
(assuming $\theta < \min\set{1,L}$
 and $0\le\psi<1$)
	\[
	\P(\<x,v\>^2\le\theta)\le
	1-\f{1-\theta}{L}
		+
		\f{2\tau^2 \exp\{ -\psi L/(2\tau^2) \} }{(L-\theta)
			\psi(1-\psi)^{1/2}}
		\,.
	\]
If we take $\psi=2/3$, $L = \beta\tau^2\ge1$,
and $\theta \le \min\set{L,1}/100 $, then
	\[\P(\<x,v\>^2\le\theta)
	\le 1-\f{1}{\beta\tau^2}
	\bigg(
	1-\theta -\f{(100/99) 3^{3/2} \tau^2}{ \exp\{\beta/3\} }
	\bigg)
	\le 1-\f{1}{ 2 \tau^2 ( 8+3\log(\tau^2)) }\,,
	\]
where the last inequality is by taking $\beta=8+3\log(\tau^2)$,
and recalling $\theta \le 1/100$.
This proves the second claim.
\end{proof}
\end{lem}

The following bound is very well known (see for instance \cite[Theorem~5.39]{MR2963170}); we include the short proof here in order to have an explicit dependence on $\tau$.

\begin{lem}\label{l:outer.product}
Suppose $a_1,\ldots,a_r$ are i.i.d.\ random vectors in $\R^d$ 
satisfying \ref{e:vector.isometric}~and~\ref{e:vector.subgaussian};
and denote $A_s\equiv a_s(a_s)^\sT$. Suppose $r$ grows polynomially in $d$. Then, with very high probability,
	\[
	\bigg\|\sum_{s\le r} A_s\bigg\|_\op
	\le r/d
	+(\log d)^{6/5}
	\max\set{\tau(r/d)^{1/2},\tau^2}
	\le
	(\log d)^{5/4}
	\max\set{ r/d,\tau^2 }\,.
	\]

\begin{proof}
Denote $x=a_s$ and consider 
$Z=xx^t-\id/d$. Recalling \eqref{e:subgaussiannorm.tail.bound}, we have $\|Z\|_\op \le 2\tau^2$ with very high probability. Write $A \preccurlyeq B$ to denote that $B-A$ is positive semidefinite. It holds for any constant $M\ge0$ that
	\[0\preccurlyeq\E[Z^2]
	= \E[\|x\|^2xx^t-\id/d^2]
	\preccurlyeq
	\E[\|x\|^2xx^t]
	\preccurlyeq
	M\E[xx^t]
	+ \E[ \Ind{\|x\|^2\ge M}
	(\|x\|^2-M)xx^t ]\,.
	\]
Taking norms (and applying the triangle inequality and Jensen's inequality) gives
	\[
	\|\E[Z^2]\|_\op
	\le M/d + \E[(\|x\|^2-M)\|x\|^2
		; \|x\|^2\ge M]
	\le 2\tau^2/d\,,
	\]
where the last inequality holds for sufficiently large $d$ by another application of \eqref{e:subgaussiannorm.tail.bound}.
Combining with the truncated Bernstein bound (Proposition~\ref{p:trunc.bernstein}) gives,
with very high probability,
	\[
	\bigg\|\sum_{s\le r}
	( A_s - \id/d)\bigg\|_\op
	\le
	(\log d)^{6/5}
	\max\set{\tau(r/d)^{1/2},\tau^2}
	\]
The claimed bound follows by using the triangle inequality.
\end{proof}
\end{lem}

\subsection{Observation of contracted tensor, diagonal component} 

The key technical step in our result is the following
estimate on $W$.

\begin{ppn}\label{p:observe.contract.tensor}
Suppose $a_1,\ldots,a_r$ are i.i.d.\ random vectors in $\R^d$  satisfying \ref{e:vector.symmetric},~\ref{e:vector.isometric}~and~\ref{e:vector.subgaussian}; and suppose $r\le d^2$.
Let $G$ and $W$ be as in \eqref{e:contract.tensor}
and \eqref{e:obs.contract.tensor}.
If $\delta^2\max\set{r,d}\ge1$, then
	\[ 
	\|W-G^\dg\|_\op
	\lesssim
	\f{\max\set{1,r/d}}
	{d^{1/2}\delta},
	\]
with high probability.

\begin{proof}
Recall the notation
$A_s \equiv a_s(a_s)^\sT$.
Write $a_{st} \equiv a_s\otimes a_t$,
and denote
 $A_{st} \equiv  a_{st}(a_{st})^\sT=A_s\otimes A_t$.
We also abbreviate $E_{ij}\equiv e_i(e_j)^\sT$
where $e_i$ denotes the $i$-th standard basis vector in $\R^d$.
For $\ell\le d$ let $I^\ell,J^\ell$ denote the $d\times d$ matrices with entries
	\[
	(I^\ell)_{ij} = I_{\ell ij},\quad
	(J^\ell)_{fg} = J_{\ell fg}\,.
	\]
Recall from \eqref{e:contract.tensor} that
	\[G= \sum_{s,t} \<a_s,a_t\> A_{st}
	= G^\dg + G^\cross\,.
	\]
The observed version $W$ of $G$
can be decomposed analogously:
	\[
	W = \sum_{s,t} 
	C_{st} \odot A_{st}
	= W^\dg+W^\cross
	\]
where, for each $s,t\le r$, we define
the $d^2\times d^2$ matrix
	\beq\label{e:matrix.of.ips}
	C_{st}
	= \f1{\delta^2}
	\sum_{\ell\le d} a_{s\ell} a_{t\ell}
	(I^\ell \otimes J^\ell)\,.
	\eeq
Let $\E''$ denote expectation over 
the indicators $I$ and $J$;
and note that $\E'' W^\dg = G^\dg$.
We show below
(Propositions~\ref{p:obs.contracted.diag}~and~\ref{p:cross.obs}) that
	\begin{align}
	\label{e:obs.contracted.bounds.diag}
	\|W^\dg-G^\dg\|_\op
	&\lesssim
	\max\bigg\{
	\f{1}{d\delta^{1/2}},
	\f{r}{d^{3/2} \delta}
	\bigg\},\\
	\label{e:obs.contracted.bounds.cross}
	\|W^\cross\|_\op
	&\lesssim
	\f{\max\set{1,r/d}}{d^{1/2}\delta}\,.
	\end{align}
Since
$W-G^\dg
=(W^\dg-G^\dg)+W^\cross$, the triangle inequality gives the claimed bound.
\end{proof}
\end{ppn}

We now prove \eqref{e:obs.contracted.bounds.diag} and \eqref{e:obs.contracted.bounds.cross}. These proofs are slightly involved, and may not offer much insight on a casual reading. We supplement these proofs with an analysis of $G^\dg$ and $G^\cross$, given in Appendix~\ref{appx:remarks}.
In particular, our analysis of $W^\cross$ is modelled after the analysis of $G^\cross$ (which is easier, and corresponds to the special case $\delta=1$). Appendix~\ref{appx:remarks}
is not needed for the proof of Theorem~\ref{t:three} but may supply some intuition. We now turn to the analysis of 
	\beq\label{e:obs.contracted.diag}
	W^\dg
	=\sum_{s\le r} C_{ss}\odot A_{ss}\,.
	\eeq

\begin{ppn}\label{p:obs.contracted.diag} Suppose $a_1,\ldots,a_r$ are i.i.d.\ random vectors in $\R^d$  satisfying \ref{e:vector.symmetric},~\ref{e:vector.isometric}~and~\ref{e:vector.subgaussian}. Let $G^\dg$ and $W^\dg$ be as in \eqref{e:rewrite.G.diag} and \eqref{e:obs.contracted.diag}.
If $\delta^2\max\set{r,d} \ge1$, then
	\[ 
	\|W^\dg-G^\dg\|_\op
	\lesssim
	\max\bigg\{
	\f{1}{d\delta^{1/2}},
	\f{r}{d^{3/2} \delta}
	\bigg\}\,,
	\]
with very high probability.

\begin{proof}
Let $a_{\max}$
denote the maximum of all the values
$|a_{s,i}|$ ($s\le r$, $i\le d$)
and $|\<a_s,a_t\>|$ ($s\ne t$); we have $a_{\max}\lesssim d^{-1/2}$ with very high probability.
Let $S^\ell \equiv 
I^\ell \otimes J^\ell
-\E''(I^\ell \otimes J^\ell)$ and
	\beq\label{e:first.repn.diag.error}
	Z^\ell
	\equiv \f{1}{\delta^2}
	\sum_{s\le r}
	(a_{s\ell})^2 S^\ell
	\odot A_{ss}
	= \f{1}{\delta^2}
	\sum_{s\le r}
	(a_{s\ell})^2
	(\diag a_{ss})
	S^\ell
	(\diag a_{ss})
	\,.
	\eeq
Under the randomness of $I$ and $J$,
the matrices $Z^\ell$ are independent with zero mean, and
	\[
	W^\dg-G^\dg
	=W^\dg-\E'' W^\dg
	=\sum_{\ell\le d} Z^\ell\,.
	\]
Note that $\|\E'' I^\ell\|_\op =
\|\delta \I\I^\sT\|_\op = d\delta$,
while the Bernstein matrix inequality 
(Proposition~\ref{p:bernstein.rect})
gives, with very high probability,
$\|I^\ell-\E'' I^\ell\|_\op
\lesssim \max\set{(d\delta)^{1/2},1}$.
It follows from the triangle inequality that
$\|I^\ell\|_\op\lesssim \max\set{d\delta,1}$, and so
	\[
	\|S^\ell\|_\op
	\le
	\|(I^\ell-\E''I^\ell)\otimes J^\ell\|_\op
	+\|(\E''I^\ell)\otimes
		(J^\ell-\E''J^\ell)\|_\op
	\lesssim
	\max\set{(d\delta)^{3/2},1}\,.
	\]
Recalling the definition \eqref{e:first.repn.diag.error}, we conclude
that with very high probability
	\beq\label{e:Z.ell.ess.sup}
	\|Z^\ell\|_\op
	\lesssim
	\f{r(a_{\max})^6}{\delta^2}
	\|S^\ell\|_\op
	\lesssim
	\max\bigg\{
		\f{r}{d^{3/2}\delta^{1/2}},
		\f{r}{d^3\delta^2}
		\bigg\}
	\eeq
Next note that we can express
$Z^\ell$ as $\delta^{-2} S^\ell \odot T^\ell$ where
	\[T^\ell\equiv
	\sum_{s\le r} (a_{s,\ell})^2
	A_{ss}\,.
	\]
The $d^2\times d^2$ matrix $T^\ell$ is symmetric, and satisfies
the entrywise bound 
$\|T^\ell\|_\infty
\le r(a_{\max})^6
\lesssim r/d^3$. Writing $M^\ell \equiv (T^\ell)^2$, we compute
	\[
	\Sigma^\ell
	=\E''[Z^\ell(Z^\ell)^\sT]
	= \underbrace
	{\f{(1-\delta)^2}{\delta^2}
	\Proj(M^\ell)}_{\Sigma^{\ell,0}}
	+ \underbrace{
	\f{1-\delta}{\delta}
	M^\ell
	\odot ( \id_d \otimes \I\I^\sT )
	}_{\Sigma^{\ell,1}}
	+ \underbrace{
	\f{1-\delta}{\delta}
	M^\ell
	\odot ( \I\I^\sT \otimes \id_d )
	}_{\Sigma^{\ell,2}}\,.
	\]
Each $\Sigma^{\ell,0}$ is simply a diagonal matrix, so
	\[
	\|\Sigma^{\ell,0}\|_\op
	\le \f{\|M^\ell\|_\infty}{\delta^2}
	\le \f{d^2(\|T^\ell\|_\infty)^2}{\delta^2}
	\lesssim
	\f{r^2}{d^4\delta^2}\,.
	\]
Let $M^{(i)\ell}$ denote the $d\times d$ matrix with entries
$(M^{(i)\ell})_{fg}
= (M^\ell)_{if,ig}$.
We can decompose
$M^{(i)\ell}$ as the sum of two components,
	\begin{align*}
	M^{(i)\ell,\dg}
	&= \sum_{s\le r}
	(a_{s\ell} )^4 
	\|a_s\|^4
	(a_{s,i})^2
	A_s\,,\\
	M^{(i)\ell,\cross}
	&=
	\sum_{s\ne t}
	(a_{s\ell} a_{t\ell})^2
	\<a_s,a_t\>^2
	a_{s,i} a_{t,i}
	a_s(a_t)^\sT\,.
	\end{align*}
We have $\|M^{(i)\ell,\cross}\|_\op\le r^2 d (a_{\max})^{10}\lesssim r^2/d^4$, 
while 
    \[|M^{(i)\ell,\dg}\|_\op
    \lesssim d^{-3}\|\sum_{s\le r} A_s\|_{\op}
    \lesssim \max\set{1,r/d}/d^3\]
using Lemma~\ref{l:outer.product}.
Combining gives
$\|M^{(i)\ell}\|_\op
\lesssim \max\set{1,r^2/d }/d^3$, and so
	\[
	\|\Sigma^{\ell,1}\|_\op
	\le
	\f{1}{\delta}
	\bigg\|
	\sum_{i\le d} E_{ii} \otimes M^{(i)\ell}
	\bigg\|_\op
	\lesssim \f{\max\set{1,r^2/d }}
	{d^3\delta}\,.
	\]
It follows from the above estimates that
	\[\bigg\|
	\sum_{\ell\le d} \Sigma^\ell
	\bigg\|_\op^{1/2}
	\lesssim 
	\max\bigg\{
	\f{r}{d^{3/2}\delta},
	\f{1}{d\delta^{1/2}},
	\f{r}{d^{3/2}\delta^{1/2}}
	\bigg\}
	=\max\bigg\{
	\f{r}{d^{3/2}\delta},
	\f{1}{d\delta^{1/2}}
	\bigg\}
	\,.
	\]
Combining with \eqref{e:Z.ell.ess.sup}
and the truncated Bernstein bound
(Proposition~\ref{p:trunc.bernstein}) gives
	\[
	\|W^\dg-G^\dg\|_\op
	\lesssim
	\max\bigg\{
	\f{1}{d\delta^{1/2}},
	\f{r}{d^{3/2} \delta},
	\f{r}{d^3\delta^2}
	\bigg\}\,.
	\]
It follows from our assumptions that
	 \beq\label{e:triv.lbd}
	 d^3\delta^2
	\ge \delta^2\max\set{r,d} \ge1\,,
	\eeq
and the claim follows.
\end{proof}
\end{ppn}

\subsection{Observation of contracted tensor, cross component} We now turn to analyzing
	\beq\label{e:obs.contracted.tensor.cross}
	W^\cross
	= \sum_{s\ne t} C_{st}
	\odot A_{st}\,,
	\eeq
where $C_{st}$ is as in \eqref{e:matrix.of.ips}
and $A_{st}\equiv A_s\otimes A_t$.

\begin{ppn}\label{p:cross.obs} 
Suppose $a_1,\ldots,a_r$ are i.i.d.\ random vectors in $\R^d$  satisfying \ref{e:vector.symmetric},~\ref{e:vector.isometric}~and~\ref{e:vector.subgaussian}; and suppose $r\le d^2$
and $\delta^2\max\set{r,d}\ge1$. Then the matrix  $W^\cross$ of \eqref{e:obs.contracted.tensor.cross}
satisfies
	\[
	\|W^\cross\|_\op \lesssim
	\f{\max\set{1,r/d}}{d^{1/2}\delta}
	\]
with very high probability.

\begin{proof}
By the symmetry assumption~\ref{e:vector.symmetric} and the
matrix decoupling inequality
(Proposition~\ref{p:decoupling}), 
it suffices to prove the bound of Proposition~\ref{p:cross.obs}
for
	\[
	W^\sign
	=\sum_{s\ne t}
	\sgns_s\sgnt_t
	C_{st}\odot A_{st}
	\]
in place of $W^\cross$.
Recalling the notation
$E_{ij}\equiv e_i(e_j)^\sT$, we have
	\[
	C_{st}
	= \sum_{i,j\le d}
	E_{ij}\otimes C_{(ij)st}
	= \sum_{f,g\le d}
	C_{st(fg)}
	\otimes E_{fg}
	\]
where $C_{(ij)st}$ and $C_{st(fg)}$ are $d\times d$ matrices with entries
	\[(C_{(ij)st})_{fg}
	=(C_{st(fg)})_{ij}
	= (C_{st})_{if,jg}\,.\]
After some straightforward manipulations we find
	\begin{align}
	W^\sign
	&= \sum_{i,j\le d}
		(E_{ij}\otimes \I\I^\sT)
		\odot
		\overbrace{\bigg\{
		\sum_{s\ne t}\sgns_s\sgnt_t
		A_s\otimes (C_{(ij)st}\odot A_t)
		\bigg\}}^{W^{(ij)}}
	\label{e:obs.cross.decomp.one}\\
	&=\sum_{f,g\le d}
	(\I\I^\sT \otimes E_{fg})
	\odot
	\underbrace{\bigg\{
	\sum_{s\ne t}\sgns_s\sgnt_t
	(C_{st(fg)} \odot A_s)\otimes A_t
	\bigg\}}_{W^{fg}}.
	\label{e:obs.cross.decomp.two}
	\end{align}
We will show below 
(Lemma~\ref{l:obs.cross.bound}) that
	\[
	\|W^{(ij)}\|_\op
	\lesssim
	\f{\max\set{1,r/d}}{d^{1/2}\delta}
	\]
with very high probability. Let $\E'$ denote expectation over $I$ only; we then have $\E'W^{(ij)}=\E' W^\sign$. 
Under the assumptions
$r\le d^2$ and 
$\delta^2\max\set{r,d}\ge1$,
we show below (Lemmas~\ref{l:obs.cross.bound}~and~\ref{l:cross.obs.error}) that
	\[\max\bigg\{
	\|\E' W^\sign\|_\op,
	\|W^\sign-\E'W^\sign\|_\op
	\bigg\}
	\lesssim
	\f{\max\set{1,r/d}}{d^{1/2}\delta}\]
with very high probability.
The claimed bound follows from the triangle inequality.
\end{proof}
\end{ppn}

\begin{lem}\label{l:obs.cross.bound}
In the setting of Proposition~\ref{p:cross.obs},
the matrix $W^{(ij)}$ of \eqref{e:obs.cross.decomp.one} satisfies
	\[\|W^{(ij)}\|_\op
	\lesssim
	\f{\max\set{1,r/d}}{d^{1/2}\delta}
	\]
with very high probability.

\begin{proof}
Fix $i,j$ and abbreviate
$\Gamma_{st} \equiv C_{(ij)st}$,
so $\Gamma_{st}$ is a $d\times d$ matrix with entries
	\[(\Gamma_{st})_{fg}
	=\f1{\delta^2}
	\sum_{\ell\le d}
	I_{\ell ij}J_{\ell fg}
	a_{s\ell}a_{t\ell}\,.
	\]
It follows from the standard Bernstein inequality that, with very high probability,
	\[
	\|\Gamma_{st}\|_\infty
	\lesssim \max\bigg\{
	\f{1}{d^{1/2}\delta},
	\f{1}{d\delta^2}
	\bigg\}\,.
	\]
Now denote
$W_{st}\equiv \Gamma_{st} \odot A_t$,
and note that
	\[
	W^{(ij)}
	= \sum_{s\le r}\sgns_s A_s
	\otimes\bigg(
	\sum_{t\in[r]\setminus s}
	\sgnt_t W_{st}
	\bigg)\,.
	\]
We can express
$W_{st} = (\diag a_t)\Gamma_{st}(\diag a_t)$, so,
with very high probability,
	\beq\label{e:y.st.max.bound}
	\|W_{st}\|_\op
	\le (\|a_t\|_\infty)^2
	\|\Gamma_{st}\|_\op
	\lesssim d^{-1} \|\Gamma_{st}\|_\op
	\le \|\Gamma_{st}\|_\infty
	\lesssim
	\max\bigg\{
	\f{1}{d^{1/2}\delta},
	\f{1}{d\delta^2}
	\bigg\}\,.
	\eeq
Conditional on $a_s,I,J$, then the  $W_{st}$ (indexed by $t\in[r]\setminus s$) are independent. For $f,g\le d$ we have
	\[(W_{st}(W_{st})^\sT)_{fg}
	=\f{1}{\delta^4} a_{tf} a_{tg}
	\sum_{k\le d}(a_{tk})^2
	\bigg(\sum_{u\le d}
	I_{uij} J_{ufk} a_{su}a_{tu}
	\bigg)
	\bigg(\sum_{v\le d}
	I_{vij} J_{vgk} a_{sv}a_{tv}
	\bigg)\,.\]
Let $\E_s$ denote expectation conditional on $a_s,I,J$; we now estimate $\Sigma_{st}\equiv\E_s[W_{st}(W_{st})^\sT]$, making use of the symmetry assumption~\ref{e:vector.symmetric}.
On the diagonal ($f=g$),
only the $u=v$ terms survive, so
	\[
	(\Sigma_{st})_{ff}
	= \f{1}{\delta^4 }\sum_{k,u\le d}
	(a_{su})^2
	I_{uij} J_{ufk}
	\E[(a_{tf} a_{tk} a_{tu})^2]
	\lesssim
	\max\bigg\{
	\f{1}{(d\delta)^2},
	\f{1}{(d\delta)^4}
	\bigg\}\,,
	\]
where the last bound holds with very high probability over $a_s,I,J$.
Off the diagonal ($f\ne g$) we must have
 $\set{u,v}=\set{f,g}$, so
	\[
	|(\Sigma_{st})_{fg}|
	=\bigg| \f{1}{\delta^4}
	a_{sf} a_{sg}
	I_{fij} I_{gij}
	\sum_{k\le d}
	(J_{ffk}J_{ggk}+J_{gfk}J_{fgk})
	\E[(a_{tf}a_{tg}a_{tk})^2]
	\bigg|\lesssim
	I_{fij} I_{gij}
	\max\bigg\{
	\f1{d^3\delta^2},
	\f{1}{(d\delta)^4}
	\bigg\}
	\]
where the last bound holds with very high probability over $a_s,J$. 
Then, with very high probability over $I$,
the number of non-zero entries in
$\Sigma_{st}$
is $\lesssim \max\set{1,(d\delta)^2}$, so
	\[\|\Proj_\perp(\Sigma_{st})\|_\F
	\lesssim 
	\max \bigg\{
	\f{1}{d^2\delta},
	\f{1}{(d\delta)^4},
	\f{1}{(d\delta)^3},
	\bigg\}
	\,.\]
Combining the diagonal and off-diagonal estimates gives altogether
	\[
	\bigg\|
	\sum_{t\in[r]\setminus s}
	\Sigma_{st}\bigg\|_\op
	\lesssim 
	\max\bigg\{
	\f{r}{(d\delta)^2},
	\f{r}{(d\delta)^4}
	\bigg\}\,.
	\]
Combining with \eqref{e:y.st.max.bound} 
and the truncated Bernstein bound
(Proposition~\ref{p:trunc.bernstein}) gives
	\[
	\bigg\|
	\sum_{t\in[r]\setminus s}
	\sgnt_t
	\Gamma_{st}\odot A_t
	\bigg\|_\op^2
	\lesssim 
	\max\bigg\{
	\f{1}{d\delta^2},
	\f{r}{(d\delta)^2},
	\f{1}{d^2\delta^4},
	\f{r}{(d\delta)^4}
	\bigg\}\,.
	\]
It then follows from the matrix Rademacher bound
(Proposition~\ref{p:rademacher})
that
	\[
	\|W^{(ij)}\|_\op^2\lesssim
	\bigg(\max_{s\le r}
	\bigg\|
	\sum_{t\in[r]\setminus s}
	\sgnt_t
	\Gamma_{st}\odot A_t
	\bigg\|_\op^2\bigg)
	\bigg\|\sum_{s\le r} A_s\bigg\|_\op\,.
	\]
with very high probability. Combining with Lemma~\ref{l:outer.product}
gives
	\[
	\|W^{(ij)}\|_\op^2\lesssim
	\f{1}{d\delta^2}
	\max\bigg\{ \bigg(
	\max\bigg\{1, \f{r}{d}\bigg\}
	\cdot
	\max\bigg\{
		1,\f{r}{d},\f{1}{d\delta^2}
		\bigg\}\bigg),
	\f{r^2}{d^4\delta^2}
	\bigg\}\,.\]
The claimed bound then follows using the assumptions $r\le d^2$ and $\delta^2\max\set{r,d}\ge1$.
\end{proof}
\end{lem}

\begin{lem}\label{l:cross.obs.error}
In the setting of Proposition~\ref{p:cross.obs},
with $W^\sign$ as in \eqref{e:obs.cross.decomp.one}
and $\E'$ denoting expectation
over $I$ only, 
the matrix $Z=W^\sign-\E'W^\sign$
satisfies
	\[
	\|Z\|_\op\lesssim
	\f{\max\set{1,r/d}}{d^{1/2}\delta}
	\]
with very high probability.

\begin{proof}
Recalling \eqref{e:obs.cross.decomp.two},
we can further decompose
	\[Z^{fg} = W^{fg}-\E'W^{fg}
	=\sum_{\ell\le d}
	Z^{\ell fg}
	\]
where $Z^{\ell fg}$ is defined as
	\[
	Z^{\ell fg}
	\equiv \left.
	\underbrace{\bigg\{
	\f{1}{\delta^2}
	J_{\ell fg}
	(I^\ell-\delta\I\I^\sT)
	\odot
	\overbrace{\bigg(
	\sum_{s\ne t}
	\sgns_s\sgnt_t
	a_{s\ell}a_{t\ell}
	a_{tf}a_{tg} A_s
	\bigg)
	\bigg\}}^{T^{\ell fg}}
	}_{M^{\ell fg}} \right. \otimes E_{fg}
	\]
Recalling that $\E'$ denotes expectation over $I$ only, we have
	\[\E'[M^{\ell fg}(M^{\ell fg})^\sT]
	=
	\f{ J_{\ell fg}(1-\delta)}{\delta^3}
	\Proj\Big( T^{\ell fg}
		(T^{\ell fg})^\sT \Big)\,.\]
By a Chernoff bound, we have 
$\|T^{\ell fg}\|_\infty\lesssim r/d^3$ with very high probability, so
	\[
	\|\E'[M^{\ell fg}(M^{\ell fg})^\sT]\|_\op
	\le
	J_{\ell fg}
	\f{d\|T^{\ell fg}\|_\infty^2}{\delta^3}
	\lesssim
	J_{\ell fg}
	\f{r^2}{d^5\delta^3}\,.
	\]
Next note that
	$Z^{\ell fg}(Z^{\ell fg})^\sT
	=
	(M^{\ell fg}(M^{\ell fg})^\sT)
	\otimes
	E_{ff}$, so altogether
	\[\bigg\|
	\sum_{f,g\le d}
	\E'[
	Z^{fg}(Z^{fg})^\sT]
	\bigg\|_\op
	\le
	\max_{f\le d}\bigg\{
	\sum_{\ell,g\le d}
	\|\E'[M^{\ell fg}(M^{\ell fg})^\sT]
	\|_\op
	\bigg\}
	\lesssim 
	\f{r^2
	\max\set{ d^2\delta, 1}}{d^5\delta^3}
	= \f{r^2}{d^3\delta^2}
	\]
where the bound holds with very
high probability over $J$, and 
the last step uses \eqref{e:triv.lbd}. The same argument as in Lemma~\ref{l:obs.cross.bound} gives
(using  $r\le d^2$ and $\max\set{r,d}\delta^2\ge1$) 
	\[
	\|Z^{fg}\|_\op
	\lesssim 
	\f{\max\set{1,r/d}}{d^{1/2}\delta}
	\]
with very high probability. 
Combining the above estimates with the truncated matrix Bernstein inequality (Proposition~\ref{p:trunc.bernstein})
gives the claimed bound.
\end{proof}
\end{lem}

\subsection{Tensor completion algorithm}

Recall $G=G^\dg+G^\cross$ from \eqref{e:contract.tensor}, and $W=W^\dg+W^\cross$ from \eqref{e:obs.contract.tensor}. We have from Proposition~\ref{p:observe.contract.tensor} that, with very high probability,
	\beq\label{e:conclusion.bound.eps}
	\epsilon=
	\|G^\dg-W\|_\op
	\lesssim
	\f{\max\set{1,r/d}}{d^{1/2}\delta}\,.
	\eeq
Choose $\delta$ large enough such that
$\epsilon\ll\ETA\ll1$, where $\ETA$ is a parameter to be determined. Let
$P$ be the orthogonal projection onto the subspace of $(\R^d)^{\otimes 2}$
spanned by singular vectors of $G^\dg$ with singular values $\ge 2\ETA$.
Let $Q$ be the orthogonal projection onto the subspace of $(\R^d)^{\otimes 2}$ spanned by singular vectors of $W$ with singular values $\ge\ETA$.
Denote the complementary projections
as $\bar{P}
\equiv \id_{d^2}-P$
and $\bar{Q}
\equiv \id_{d^2}-Q$. It follows by Wedin's theorem
(Proposition~\ref{p:wedin})
that
	\beq\label{e:tensor.apply.wedin}
	\|P \bar{Q}\|_\op
	\le\epsilon/\ETA\ll1.
	\eeq
Recall from \eqref{e:obs.contract.tensor}
the formation of $W$ using indicators $I,J$.
Let $K$ be an independent copy of $I$, and
let $\hbT$ denote the tensor with entries
$(\hbT)_{ijk} = \delta^{-1} K_{ijk}\bT_{ijk}$.
Define the estimator
	\beq\label{e:overcomplete.estimator}
	\hbTstar
	= \hbT(\id_d\otimes Q)\,.
	\eeq
In what follows we will show that $\hbTstar$ is close to $\bT$ in Frobenius norm, where
	\[
	\|\bT\|_\F^2
	=\sum_{s\le r} \|a_s\|^6
	+ \sum_{s\ne t} \<a_s,a_t\>^3\,.
	\]
Recalling the proof of Proposition~\ref{p:contracted.diag.lbd},
we have
	\[
	\sum_{s\le r} \|a_s\|^6
	\asymp r,\quad
	\bigg|
	\sum_{s\ne t} \<a_s,a_t\>^3
	\bigg|
	\lesssim \f{r}{d^{3/2}}\,,
	\]
so altogether $\|\bT\|_\F\asymp r^{1/2}$.

\begin{lem}\label{l:tensor.svd}
Suppose $a_1,\ldots,a_r$ are i.i.d.\ random vectors in $\R^d$  satisfying \ref{e:vector.symmetric},~\ref{e:vector.isometric}~and~\ref{e:vector.subgaussian}; and suppose $r$ grows polynomially in $d$.
Let $\bT$ be as in \eqref{e:recall.three.tensor},
and let $\bar{P}=\id-P$ as above.
Then it holds with very high probability that
$\|\bT(\id_d\otimes\bar{P})\|_\F
	\lesssim
	\ETA^{1/4}r^{1/2}$.

\begin{proof}
Denote $\theta_{st}
\equiv \< \bar{P}(a_s\otimes a_s),
\bar{P}(a_t\otimes a_t)\>$. 
By definition, $\|\bar{P}G^\dg\bar{P}\|_\op\le2\ETA$, so
	\beq\label{e:PGP}
	\ETA \|a_s\otimes a_s\|^2
	\ge
	(a_s\otimes a_s)^\sT\bar{P}G^\dg
	\bar{P}(a_s\otimes a_s)
	=\sum_{t\le r}
	\|a_t\|^2  (\theta_{st})^2
	\ge \|a_s\|^2 (\theta_{ss})^2\,.
	\eeq
Let $(\sgns_s)_{s\le r}$ be a collection of symmetric random signs:
by assumption~\ref{e:vector.symmetric},
the original tensor $\bT$ is equidistributed as
	\[\bT^\textup{sgn}
	=\sum_s \sgns_s 
	a_s\otimes a_s\otimes a_s\,.\]
Note that $\bT$ and $\bT^\textup{sgn}$
map to the same $G^\dg$,
so the projection matrix $P$
is independent of the signs $\sgns_s$.
Therefore
$\|\bT(\id_d\otimes\bar{P})\|_\F^2$
is equidistributed as
	\[\|\bT^\textup{sgn}
	(I\otimes\bar{P})\|_\F^2
	=\sum_{s\le r}
	\|a_s\|^2 \theta_{ss}
	+ \sum_s\sgns_s
	\bigg(\sum_{t\in[r]\setminus s}
	\sgns_t
	\<a_s,a_t\>
	\theta_{st}\bigg)\,.\]
Recall from \eqref{e:PGP}
that $|\theta_{ss}| \le \ETA^{1/2} \|a_s\|$,
so the first term is $\lesssim\ETA^{1/2}r$.
Meanwhile, by
combining \eqref{e:PGP}
with the decoupling inequality and the
Rademacher bound,
the second term is $\lesssim\ETA^{1/2} (r/d)^{1/2}$.
The claimed bound follows.
\end{proof}
\end{lem}

\begin{proof}[Proof of Theorem~\ref{t:three.bernoulli}]
We decompose
	\[\bT-\hbTstar
	=\bT(I\otimes \bar{P}\bar{Q})
	+\bT(I\otimes P\bar{Q})
	+(\bT-\hbT)(I\otimes Q)\,.\]
Since $\bar{Q}$ is a projection matrix
we have $\|\bar{Q}\|_\op\le1$, so
	\[\|\bT(I\otimes \bar{P}\bar{Q})
	\|_\F
	\le
	\|\bT(I\otimes \bar{P})
	\|_\F
	\lesssim \ETA^{1/4} r^{1/2}
	\asymp
	\ETA^{1/4} \|\bT\|_\F
	\]
by Lemma~\ref{l:tensor.svd}.
Recall from \eqref{e:tensor.apply.wedin}
that $\|P\bar{Q}\|_\op\le\epsilon/\ETA\ll1$;
we then have
	\[
	\|\bT(I\otimes P\bar{Q})\|_\F
	\le
	\|\bT\|_\F
	\|P\bar{Q}\|_\op
	\le (\epsilon/\ETA)
	\|\bT\|_\F\,.\]
Lemma~\ref{l:rank.Q} gives $\rank Q\le r$,
and combining with Lemma~\ref{l:tensor.conc}
gives
	\[\|(\bT-\hbT)(I\otimes Q)\|_\F
	\le
	r^{1/2}
	\|\unfold^{1\times 2}\big((\bT-\hbT)(I\otimes Q)\big)\|_\op
	\lesssim
	\bigg(
	\f{r}{d^{3/2}}
	\f{\max\set{1,r/d}}{d^{1/2}\delta}
	\bigg)^{1/2}
	\|\bT\|_\F\,.
	\]
The result follows by setting
$\ETA$
equal
to the parameter $\lmstar$ of 
the theorem statement,
and then recalling the bound on $\epsilon$ from
\eqref{e:conclusion.bound.eps}.
\end{proof}

\section{Remarks on contracted tensor}
\label{appx:remarks}

This section supplements Appendix~\ref{appx:three}
by analyzing $G$ (of \eqref{e:contract.tensor}).
As noted above, the estimates below are not required for the proof of Theorem~\ref{t:three}.
We include them because they may supply some intuition, and may be useful for related problems such as tensor decomposition.

\subsection{Contracted tensor, diagonal component} We begin with the diagonal component
	\beq\label{e:rewrite.G.diag}
	G^\dg
	= \sum_{s\le r}
	\|a_s\|^2 (a_s\otimes a_s)
	(a_s\otimes a_s)^\sT\,.
	\eeq
For this component, we have a slightly better estimate if we make the additional
assumption that $\tau^2 \le 21/20$. This is due to the following

\begin{cor}\label{c:combined.lbd}
Suppose $x$ is a random vector in $\R^d$ satisfying  \ref{e:vector.isometric}~and~\ref{e:vector.subgaussian} with $\tau^2 \le 21/20$. Then it holds for any deterministic $v\in\R^d$ that
	\[
	\P( \|x\|^2 \ge 1/100
	\textup{ and }
	d\<x,v\>^2 \ge \|v\|^2/100)
	\ge 1/1000
	\]
for sufficiently large $d$.

\begin{proof}
Follows by Lemma~\ref{l:lower.tail.bound} and a union bound.
\end{proof}
\end{cor}

Applying this corollary, we obtain the following estimates for the spectral norm of $G^\dg$:

\begin{ppn}\label{p:contracted.diag.lbd}
Suppose $a_1,\ldots,a_r$ are i.i.d.\ random vectors in $\R^d$ satisfying \ref{e:vector.isometric}~and~\ref{e:vector.subgaussian};
and define $G^\dg$ as in \eqref{e:rewrite.G.diag}.
Suppose that $r$ grows at least polynomially in $d$, i.e., that $(\log r)/(\log d)$ stays bounded away from zero.
\begin{enumerate}[a.]
\item There exists an absolute constant $c$ such that, with very high probability,
	\[
	\|G^\dg\|_\op 
	\ge 1 - \f{c}{\log d}\,.
	\]
\item Suppose additionally that
\ref{e:vector.subgaussian}
is satisfied with $\tau^2\le21/20$;
and that
 $(\log r)/(\log d)$ stays bounded away from infinity as well as from zero. Then there exists an absolute constant $c$ such that, with very high probability,
 	\[
	\|G^\dg\|_\op
	\ge c\bigg( 1 +
	\f{r/d}{(\log d)^{3/5}} \bigg)\,.
	\]
\end{enumerate}

\begin{proof} For any $s$ such that $a_s\ne0$ we have
	\[ \|G^\dg\|_\op
	\ge\f{\< a_s\otimes a_s,
	G^\dg(a_s\otimes a_s)\>}
	{ \|a_s\otimes a_s\|^2 }
	= \|a_s\|^6
	+ \sum_{t\in[r]\setminus s}
	\f{\<a_s,a_t\>^4}{\|a_s\|^4}
	\ge \|a_s\|^6\,.
	\]
Since $r$ grows at least polynomially in $d$,
Lemma~\ref{l:lower.tail.bound} gives  $\max_s\|a_s\|\ge 1-O(1)/\log d$. This implies the result of (a). Turning to the proof of (b), we will lower bound
	\[\|G^\dg\|_\op
	\ge
	\f{\|G^\dg u\|}{\|u\|},\quad
	u = \sum_{s\le r} a_s \otimes a_s\,.
	\]
From Lemma~\ref{lem:subgaussiannorm}, if $x$ is $(\tau^2/d)$-subgaussian, then
	\beq\label{e:subgaussiannorm.tail.bound}
	\P(\|x\|^2 \ge t)
	\le \E\exp\bigg\{
		\f{d(\|x\|^2-t)}{4\tau^2}
		\bigg\}
	\le 
	\exp\bigg\{ \f{3d}{8}
		- \f{dt}{4\tau^2} \bigg\}\,,
	\eeq
so $\P(\|x\|^2\ge 2\tau^2) \le \exp\{-d/8\}$.
For any deterministic $v\in\R^d$ with $\|v\|^2=d$, we have (by the same calculation as above, for the case $d=1$)
	\beq\label{e:SubGaussianIP.tail.bound}
	\P( \<x,v\>^2 \ge t )
	\le
	\E \exp\bigg\{
	\f{\<x,v\>^2 -t }{ 4\tau^2}
	\bigg\}
	\le \exp\bigg\{
		\f38 - \f{t}{4\tau^2}
		\bigg\}\,,
	\eeq
so that $\<x,v\>^2 \le (\log d)^{6/5}$
with very high probability.
Taking a union bound over $r$
(and using that $r$ is at most polynomial in $d$), we conclude that the event
	\[\bigcap_{s\le r}
	\bigg\{
	\|a_s\|^2 \le 2\tau^2
	\textup{ and }
	\max_{t \in[r]\setminus s}
	|\<a_s,a_t\>|^2 \le 
	\f{\|a_s\|^2(\log d)^{6/5}}{d}
	\bigg\}
	\]
occurs with very high probability. Combining these gives for all $s\le r$ that
	\[
	\kappa_s
	\equiv \sum_{t\le r}\<a_s,a_t\>^2
	\le \|a_s\|^2
	\bigg(\|a_s\|^2 + 
		\f{r (\log d)^{6/5}}{d}\bigg)
	\le 2\tau^2
	\bigg(2\tau^2 + 
		\f{r (\log d)^{6/5}}{d}\bigg)\,,
	\]
and so we conclude
	\beq\label{e:vec.u.ubd}
	\|u\|^2
	= \sum_{s\le r} \kappa_s
	\le r\cdot 2\tau^2
	\bigg(2\tau^2 +
	 \f{r (\log d)^{6/5}}{d}\bigg)\,.
	\eeq
We next turn to lower bounding
	\beq\label{e:G.diag.u}
	\|G^\dg u\|^2
	=
	\bigg\|\sum_{s\le r} 
	\kappa_s
	\|a_s\|^2 (a_s\otimes a_s)
	\bigg\|^2
	= \sum_{s,t}
	\kappa_s
	\kappa_t
	\|a_s\|^2
	\|a_t\|^2
	\<a_s,a_t\>^2
	\,.
	\eeq
In what follows, we use $c_i$ to denote positive absolute constants. By Lemma~\ref{l:lower.tail.bound},
since $r$ grows polynomially in $d$,
the event
	\[\bigcap_{s\le r}
	\bigg\{
	|\set{s : \|a_s\|^2 \ge 1/2}|
	\ge \f{r}{3\tau^2}
	\textup{ and }
	\bigg|\bigg\{
	t\in[r]\setminus s:
	\<a_s,a_t\>^2
	\ge\f{\|a_s\|^2}{100 d}
	\bigg\}\bigg| \ge \f{r}
		{5\tau^2( 8+3\log(\tau^2) )}
	\bigg\}
	\]
occurs with very high probability.
Combining these gives
for all $s\le r$ that
	\[
	\kappa_s
	\ge
	\|a_s\|^2\bigg(
	\|a_s\|^2
	+ \f{r/d}
	{ 100 c(\tau^2) }
	\bigg)
	\ge
	\|a_s\|^2
	\bigg(
	\|a_s\|^2 + \f{c_{1} r}{d}
	\bigg)
	\,.
	\]
By Corollary~\ref{c:combined.lbd}, using the additional assumption $\tau^2\le21/20$, the event
	\[
	\bigcap_{s\le r}
	\bigg\{
	\bigg|
	\bigg\{
	t\in[r]\setminus s :
	\|a_t\|^2 \ge \f{1}{100}
	\textup{ and }
	\<a_t,a_s\>^2\ge \f{\|a_s\|^2}{100 d}
	\bigg\} \bigg|
	\ge \f{r}{2000}
	\bigg\}
	\]
also occurs with very high probability.
It follows that for all $s\le r$,
	\[
	\sum_{t\in[r]\setminus s}
	\kappa_t
	\|a_t\|^2
	\<a_s,a_t\>^2
	\ge 
	\f{c_{2}  r\|a_s\|^2 
		(1+r/d)}{d}\,,
	\]
and consequently
	\beq\label{e:G.diag.u.final.lbd}
	\|G^\dg u\|^2
	\ge
	\sum_s
	\kappa_s\|a_s\|^2
	\bigg(
	\kappa_s\|a_s\|^6
	+ \f{c_{2}  r\|a_s\|^2 
		(1+r/d)}{d}
	\bigg)
	\ge 
	c_{3}
	r (1+r/d)^3\,.\eeq
Combining \eqref{e:vec.u.ubd} and 
\eqref{e:G.diag.u.final.lbd} proves
	\[
	\|G^\dg\|_\op\ge 
	\f{c_{4}(1+r/d)}
	{(\log d)^{3/5}}\,.
	\]
Combining with the lower bound from (a)
gives the result of (b).
\end{proof}
\end{ppn}

\subsection{Contracted tensor, cross component}
Recalling \eqref{e:contract.tensor},
we now turn to showing that
	\beq\label{e:rewrite.G.cross}
	G^\cross
	= \sum_{s\ne t}
	\<a_s,a_t\> 
	(a_s\otimes a_t)
	(a_s\otimes a_t)^\sT
	\eeq
has smaller spectral norm than $G^\dg$. We follow a 
similar argument from
\cite[Propn.~5.5]{hsss}.

\begin{ppn}\label{p:Gcross}
Suppose $a_1,\ldots,a_r$ are i.i.d.\ random vectors in $\R^d$  satisfying \ref{e:vector.symmetric},~\ref{e:vector.isometric}~and~\ref{e:vector.subgaussian}; and suppose $r$ grows polynomially in $d$. Then, with very high probability,
	\[\|G^\cross\|\le 
	\f{(\log d)^4
	\tau^3}{d^{1/2}}
	\max\set{r/d,\tau^2}\,.\]

\begin{proof} Recall the notation $A_s\equiv a_s(a_s)^\sT$. Let $(\sgns_s,\sgnt_s)_{s\le r}$ be a collection of i.i.d.\ symmetric random signs. By the symmetry assumption~\ref{e:vector.symmetric}, $a_s$ is equidistributed as $\sgns_s a_s$ where the $\sgns_s$ are independent symmetric random signs, so $G^\cross$ is equidistributed as
	\[
	\sum_{s\le r}
	\sgns_s A_s
	\otimes\bigg(
	\sum_{t\in[r]\setminus s}
	\sgns_t \<a_s,a_t\> A_t\bigg)\,.
	\]
In view of the decoupling inequality  (Proposition~\ref{p:decoupling}), it is enough to prove the claimed bound for the matrix $G^\textup{sgn}$, which is defined as above but with $\sgnt_t$ in place of $\sgns_t$.
To this end, let us first bound the spectral norm of
	\[
	G_s
	\equiv \sum_{t\in[r]\setminus s}
	\sgnt_t \<a_s,a_t\> A_t\,.
	\]
Conditional on $a_s$, the summands
$G_{st}\equiv \sgnt_t \<a_s,a_t\> A_t$
are independent with zero mean.
Recalling
\eqref{e:subgaussiannorm.tail.bound}
and \eqref{e:SubGaussianIP.tail.bound},
conditional on $a_s$ it holds with very high probability that
	\[
	\|G_{st}\|_\op
	= |\<a_s,a_t\>| \|a_t\|^2
	\le \f{(\log d)^{3/5}
	\tau^2\|a_s\|}{ d^{1/2}}\,.
	\]
Next, arguing similarly as in the proof of Lemma~\ref{l:outer.product}, we have
	\[
	\|\E[(G_{st})^2 |a_s]\|_\op
	\le
	\f{(\log d)^{6/5}
	\tau^2\|a_s\|^2}{d^2}\,.
	\]
It follows using the truncated Bernstein bound
(Proposition~\ref{p:trunc.bernstein}) that,
with very high probability,
	\[
	\|G_s\|_\op
	\le
	\f{(\log d)^2 \tau \|a_s\|}{d^{1/2}}
	\max\set{(r/d)^{1/2},\tau}
	\]
for all $s\le r$.  It also holds with very high probability that $\max_s\|a_s\|^2 \le 2\tau^2$. Now consider
	\[
	G^\textup{sgn}
	= \sum_{s\le r} \sgns_s
	A_s\otimes G_s
	\]
--- recalling the matrix Rademacher bound (Proposition~\ref{p:rademacher}), we shall bound
	\[
	\sigma(G^\textup{sgn})
	=\bigg\|\sum_{s\le r} (A_s \otimes G_s)
	(A_s\otimes G_s)^\sT\bigg\|_\op^{1/2}\,.\]
Each $(A_s \otimes G_s)(A_s\otimes G_s)^\sT$
is positive semidefinite, so
	\[
	\sigma(G^\textup{sgn})
	\le
	(\max_s\|G_s\|)
	\bigg\|\sum_{s\le r} A_s(A_s)^\sT
	\bigg\|_\op^{1/2}
	\le
	\Big(\max_{s\le r}\|G_s\|\Big)
	\Big(\max_{s\le r}\|a_s\|\Big)
	\bigg\|\sum_{s\le r} A_s\bigg\|_\op^{1/2}\,.
	\]
By the preceding estimates together with Lemma~\ref{l:outer.product},
	\[
	\sigma(G^\textup{sgn})
	\le
	\f{(\log d)^{3-1/4}
	\tau^3}{d^{1/2}}
	\max\set{r/d,\tau^2}
	\]
with very high probability.
The claimed result follows by conditioning on the event that the above bound holds, and then applying Proposition~\ref{p:rademacher}.
\end{proof}
\end{ppn}

\end{document}